\documentclass[usenatbib]{mnras}

\usepackage{amssymb}

\usepackage{newtxtext,newtxmath}
\usepackage[T1]{fontenc}

\DeclareRobustCommand{\VAN}[3]{#2}
\let\VANthebibliography\thebibliography
\def\thebibliography{\DeclareRobustCommand{\VAN}[3]{##3}\VANthebibliography}

\usepackage{graphicx}	% Including figure files
\usepackage{amsmath}	% Advanced maths commands
\usepackage{color}
\usepackage{threeparttable, tablefootnote}  % table footnotes
\usepackage{stfloats}

\newcommand{\Rtwohc}{R_{\rm 200c}}
\newcommand{\Mtwohc}{M_{\rm 200c}}
\newcommand{\Rtwohm}{R_{\rm 200m}}
\newcommand{\Mtwohm}{M_{\rm 200m}}
\newcommand{\Rfivehc}{R_{\rm 500c}}

\newcommand{\eg}{{\sl e.g.}, }

\newcommand{\mtwoh}{M_{\rm 200c}}

\newcommand{\Mr}{M_{\rm r}}

\newcommand{\kms}{{\rm \, km~s}\ensuremath{^{-1}}}
\newcommand{\msol}{\ensuremath{\, {\rm M}_\odot}}    
\newcommand{\msun}{\ensuremath{\, {\rm M}_\odot}} 
         
\newcommand{\mpc}{\ensuremath{\, {\rm Mpc}}}         
\newcommand{\gpc}{\ensuremath{\, {\rm Gpc}}}

\newcommand{\Mstarsat}{\ensuremath{M_{\rm \star,\,sat}}}

\newcommand{\sigmaSat}{\ensuremath{\sigma_{\rm sat,\,1D}}}

\newcommand{\sigmaDM}{\ensuremath{\sigma_{\rm DM}}}

\definecolor{bleudefrance}{rgb}{0.19, 0.55, 0.91}
\definecolor{purple}{RGB}{128, 0, 128}

\newcommand*{\vcenteredhbox}[1]{\begingroup
\setbox0=\hbox{#1}\parbox{\wd0}{\box0}\endgroup}

\title[Calibrating the galaxy velocity bias]{Galaxy Velocity Bias in Cosmological Simulations: Towards Percent-level Calibration}

\author[Dhayaa Anbajagane]{Dhayaa Anbajagane (\vcenteredhbox{\includegraphics[height=1.2\fontcharht\font`\B]{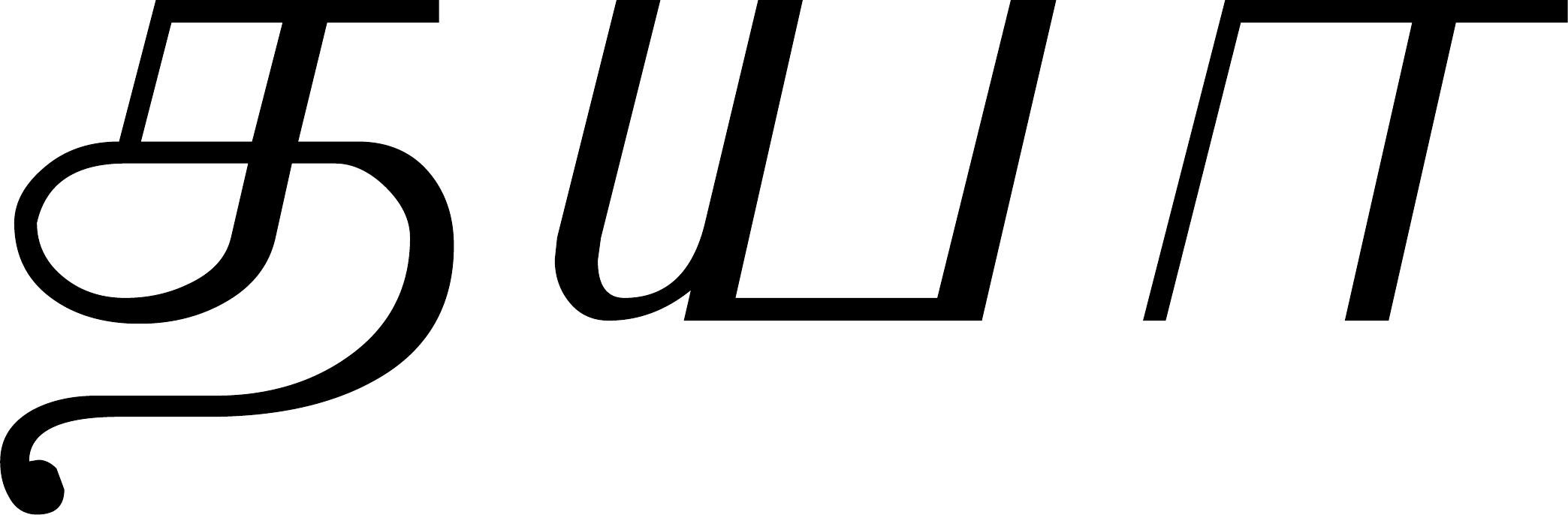}}),$^{1,\, 2}$\thanks{Corresponding author email: dhayaa@uchicago.edu}
Han Aung,$^{3}$
August E. Evrard,$^{2,\, 4}$
Arya Farahi,$^{5,\, 6}$\newauthor
Daisuke Nagai,$^{3,\, 7}$
David J. Barnes,$^8$
Weiguang Cui,$^{9}$
Klaus Dolag,$^{10,\, 11}$ 
Ian G. McCarthy,$^{12}$\newauthor
Elena Rasia,$^{13,\, 14}$
Gustavo Yepes$^{15}$
\\
\\
$^1$Department of Astronomy and Astrophysics, University of Chicago, 5640 S. Ellis Ave, Chicago, IL 60637, USA\\
$^2$Department of Physics and Leinweber Center for Theoretical Physics, University of Michigan, Ann Arbor, MI 48109, USA\\
$^3$Department of Physics, Yale University, New Haven, CT 06520, USA\\
$^4$Department of Astronomy, University of Michigan, Ann Arbor, MI 48109, USA\\
$^5$ Michigan Institute for Data Science, University of Michigan, Ann Arbor, MI 48109, USA\\ 
$^6$ Department of Statistics and
Data Science, The University of Texas at Austin, TX 78712, USA\\
$^7$ Department of Astronomy, Yale University, New Haven, CT 06520, USA\\
$^8$ Kavli Institute for Astrophysics and Space Research, Massachusetts Institute of Technology, Cambridge, MA 02139, USA\\
$^{9}$Institute for Astronomy, University of Edinburgh, Edinburgh, EH9 3HJ, UK\\
$^{10}$Max-Planck Institut f\"ur Astrophysik, Karl-Schwarzschild Str. 1, D-85741 Garching, Germany\\
$^{11}$University Observatory Munich, Scheinerstr. 1, 81679 M\"unchen, Germany\\
$^{12}$Astrophysics Research Institute, Liverpool John Moores University, 146 Brownlow Hill, Liverpool L3 5RF, UK\\
$^{13}$INAF, Osservatorio Astronomico di Trieste, via Tiepolo 11, I-34131, Trieste, Italy\\
$^{14}$IFPU - Institute for Fundamental Physics of the Universe, Via Beirut 2, 34014 Trieste, Italy\\
$^{15}$Departamento de F\'isica Te\'orica  M-8  and CIAFF,   Facultad de Ciencias,  Universidad Aut\'onoma de Madrid, E-28049 Madrid, Spain
} 

\date{Accepted XXX. Received YYY; in original form ZZZ}

\pubyear{2021}

\begin{document}
\label{firstpage}
\pagerange{\pageref{firstpage}--\pageref{lastpage}}
\maketitle

%Abstract of the paper
\begin{abstract}

Galaxy cluster masses, rich with cosmological information, can be estimated from internal dark matter (DM) velocity dispersions, which in turn can be observationally inferred from satellite galaxy velocities. However, galaxies are biased tracers of the DM, and the bias can vary over host halo and galaxy properties as well as time. We precisely calibrate the velocity bias, $b_v$ --- defined as the ratio of galaxy and DM velocity dispersions ---
as a function of redshift, host halo mass, and galaxy stellar mass threshold ($\Mstarsat$), for massive haloes ($\Mtwohc > 10^{13.5} \msun$) from five cosmological simulations: IllustrisTNG, Magneticum, Bahamas + Macsis, The Three Hundred Project, and MultiDark Planck-2.
We first compare scaling relations for galaxy and DM velocity dispersion across simulations; the former is estimated using a new ensemble velocity likelihood method that is unbiased for low galaxy counts per halo, while the latter uses a local linear regression.
The simulations show consistent trends of $b_v$ increasing with $\Mtwohc$ and decreasing with redshift and $\Mstarsat$. The ensemble-estimated theoretical uncertainty in $b_v$ is 2-3\%, but becomes \textit{percent-level} when considering only the three highest resolution simulations.
We update the mass--richness normalization for an SDSS redMaPPer cluster sample, and find our improved $b_v$ estimates reduce the normalization uncertainty from 22\% to 8\%, demonstrating that dynamical mass estimation is competitive with weak lensing mass estimation. We discuss necessary steps for further improving this precision. Our estimates for $b_v(\Mtwohc, \Mstarsat, z)$ are made publicly available.
\end{abstract}

%Select between one and six entries from the list of approved keywords.
%Don't make up new ones.
\begin{keywords}
galaxies: clusters: general -- galaxies: kinematics and dynamics -- methods: statistical
\end{keywords}

%%%%%%%%%%%%%%%%%%%%%%%%%%%%%%%%%%%%%%%%%%%%%%%%%%

%%%%%%%%%%%%%%%%% BODY OF PAPER %%%%%%%%%%%%%%%%%%
\section{Introduction}

Galaxy clusters, and their associated massive dark matter (DM) haloes, contain rich information on the composition and evolutionary history of our Universe.  This information can be extracted by connecting the observable properties of clusters to the halo mass function (HMF) --- a number density of haloes as a function of halo mass --- which then translates into constraints on cosmological parameters. A key component of this inference process is constructing a probabilistic mapping function between cluster observables and the mass of the underlying massive halo. 
Observable properties with established and understood connections are commonly referred to as halo mass proxies, and there exist many across multiple wavelengths \citep[see][for reviews]{Allen2011ClusterReview, Pratt2019ClusterMassReview}. In this work we focus on connecting satellite galaxy kinematics, measured via spectroscopy, to the host halo mass.

The collisionless DM velocity field within a halo is driven by the halo's gravitational potential.  When virial equilibrium is satisfied, meaning the average kinetic energy is half the magnitude of the potential energy, an estimate for the total halo mass can be obtained by inferring the average kinetic energy from the satellite galaxy kinematics of a halo.  \citet{Zwicky1937Virial} famously used this approach to estimate a halo mass and argue for the existence of dark matter. Note that this technique of halo mass estimation implicitly assumes that the satellite galaxies fairly trace the DM velocity field. Galaxies, however, are known to be biased tracers of the underlying DM density field \citep{Kaiser1984Bias, Davis1985EvolutionOfLSS, Bardeen1986PeakStatistics} and of the DM velocity field, as shown by the many works that we detail below. The bias pertaining to the latter case is commonly denoted the velocity bias, $b_v$, and the focus of this work is the mean $b_v$ of the halo population as a function of halo and galaxy properties.

The velocity bias is linked closely to the physics of galaxy formation, and in particular to when galaxies fall into a DM halo and become subject to dynamical friction, tidal disruption, and other non-linear effects \citep{Carlberg1990velocitybias,Carlberg1991dynamicalBias,Colafrancesco1995}. Notably, the uncertainty in this bias is the dominant systematic uncertainty in dynamical mass estimation techniques \citep[henceforth F16]{Sifon2016VelDispACT, Farahi2016StackedSpectro}; the study of F16 showed that the precision in halo mass is limited to $\approx 25\%$ due to poor knowledge of $b_v$.

Previous works have studied the qualitative and/or quantitive trends of the satellite galaxy velocity bias as a function of galaxy properties (such as stellar mass, galaxy luminosity, and redshift) using simulations \citep{Biviano2006DynamicalMass, Lau2010BaryonDissipationVelDisp, Wu2013VirialScalingPlusBias, Old2013VelBias, Munari2013VelDisp, Ye2017IllustrisVelBias, Armitage2018CEagleVelBias, Ferragamo2020VelDispEstimators} as well as observational data \citep{Biviano1992VelocitySegregation, Stein1997BrightToCool, Adami1998BrighterToCooler, Adami2000RedshiftEvol, Girardi2003VelocitySegregation, Goto2005BrighterToCooler, Barsanti2016VelocitySegregation, Nascimento2017VelocitySegregation, Bayliss2017VelDisp}. The results of these works are all consistent with brighter galaxies being kinematically cooler --- and thus having a \textit{lower} velocity bias --- than fainter galaxies.

However, \citet[][hereafter G15]{Guo2015RsdLuminosityDependance}, who used measurements of the small-scale redshift-space distortions (RSD) to infer a velocity bias as a function of galaxy luminosity, find that brighter galaxies have a \textit{higher} (not lower) velocity bias than fainter galaxies. Their estimates are thus in tension with the aforementioned studies. For example, \citet{Bayliss2017VelDisp} use observed spectra from nearly 3000 satellite galaxies in 89 clusters identified via the Sunyaev-Zel'dovich effect, and find that the velocity dispersion --- which is the second moment of a velocity field --- for brighter galaxies is $11 \pm 4$ percent lower than the velocity dispersion for the full galaxy population. It is speculated that the discrepancy between G15 and the other works arises because G15 uses all galaxies in the survey volume, and not just satellite galaxies hosted in massive haloes \citep[][see conclusions]{Ye2017IllustrisVelBias}. Another difference is that G15 use both the one-halo \textit{and} two-halo components of the velocity fields in their RSD analysis, whereas all the cluster-focused studies mentioned above limit themselves to the one-halo component alone.

Given the discrepancy in the G15 result, we require an alternative calibration for the satellite galaxy velocity bias as a function of relevant galaxy/halo properties. The other observational works noted above have studied the \textit{relative} trends of the velocity bias with galaxy luminosity but did not estimate the actual values of $b_v$. Some simulation studies have estimated both $b_v$ and its dependence on galaxy luminosity, but using alternative methodologies to that used in our work: \citet{Lau2010BaryonDissipationVelDisp, Wu2013VirialScalingPlusBias} selected the top $N$ galaxies per halo according to $\Mstarsat$ and studied the response of the velocity bias to varying $N$, and \citet{Ferragamo2020VelDispEstimators} selected the top $N\%$ of all satellite galaxies in cluster-scale haloes. While these works have shed light on the velocity bias of galaxies within host haloes, they do not provide a function or mapping for the velocity bias given a galaxy stellar mass threshold or galaxy magnitude threshold.

The two simulation-based works that \textit{have} estimated this mapping \citep{Ye2017IllustrisVelBias, Armitage2018CEagleVelBias} are limited in either using a small sample size of only the most massive haloes, or using a single simulation model. 
The former leads to larger statistical uncertainties in the bias estimates, in addition to being limited to a narrow mass range, whereas the latter cannot quantify the theoretical uncertainty in the velocity bias, i.e. the variation in the velocity bias due to different astrophysical and numerical treatments.  

In this work, we use an ensemble of simulations to calibrate the satellite galaxy velocity bias, \textit{including} the relevant theoretical uncertainty, as a function of the galaxy stellar mass threshold, host halo mass, and redshift. We extend on the previous body of work in three different directions: (i) We propose and validate a new likelihood-based estimator for the scaling parameters (normalization, slope and population intrinsic scatter) of the galaxy velocity dispersion with halo mass in the regime of low galaxy counts per halo, (ii) We perform a convergence study of the velocity bias, as well as galaxy and DM velocity dispersions, across a suite of cosmological, hydrodynamics simulations, and also an N-body simulation with galaxies painted on using a semi-analytical model, and; (iii) Finally, using predictions from the ensemble of simulations, we construct a theoretical prior on the velocity bias that incorporates the \textit{modeling uncertainty} associated with varying numerical and galaxy formation treatments. We then use this prior to refine the mean halo mass estimates previously derived in F16 for SDSS redMaPPer galaxy clusters. 

A part of our convergence study focuses on the mass-dependent population statistics --- mean and intrinsic scatter --- of DM velocity dispersion, and is thus a hydrodynamical counterpart to the original study of \citet[henceforth E08]{Evrard2008VirialScaling}, who used a large ensemble of mostly N-body simulations to set precise constraints on these quantities. Note also that we previously employed a subset of the simulations used in this work to perform similar convergence tests of mass-dependent population statistics for central and satellite galaxy properties of cluster-scale haloes; \citet{Anbajagane2020StellarStatistics} find that the distribution of residuals about the property mean relations share similar functional forms, and that the mean relations themselves have moderate offsets between simulations.

This paper is organized as follows: in \S\ref{sec:Data} we describe our simulation ensemble, and in \S\ref{sec:Methods} we define the relevant halo properties and detail the scaling relation estimators, including the aforementioned ensemble velocity likelihood method. Our results for the galaxy/DM velocity dispersion and the velocity bias are presented in \S\ref{sec:Results}, while the impact of our work for dynamical mass estimation is both demonstrated and discussed in \S\ref{sec:Cluster_Applications}. Finally, we conclude in \S\ref{sec:conclusions}. Our appendices contain results on resolution effects (Appendix \ref{appx:Res_test}), additional validation tests of the likelihood model (Appendix~\ref{appx:likelihood_tests}), and the impact of radial aperture choices on the velocity bias (Appendix~\ref{appx:Aperture}).

Throughout this work we use a spherical overdensity definition of halo mass, $M_{\rm \Delta} = \rho_\Delta [\frac{4\pi}{3} R_{\rm \Delta}^3]$,
with contrast value, $\rho_\Delta = 200\rho_c(z)$, where $\rho_c(z)$ is the critical density at redshift $z$.

\section{Data} \label{sec:Data}

\begin{figure}
    \centering
    \includegraphics[width = \columnwidth]{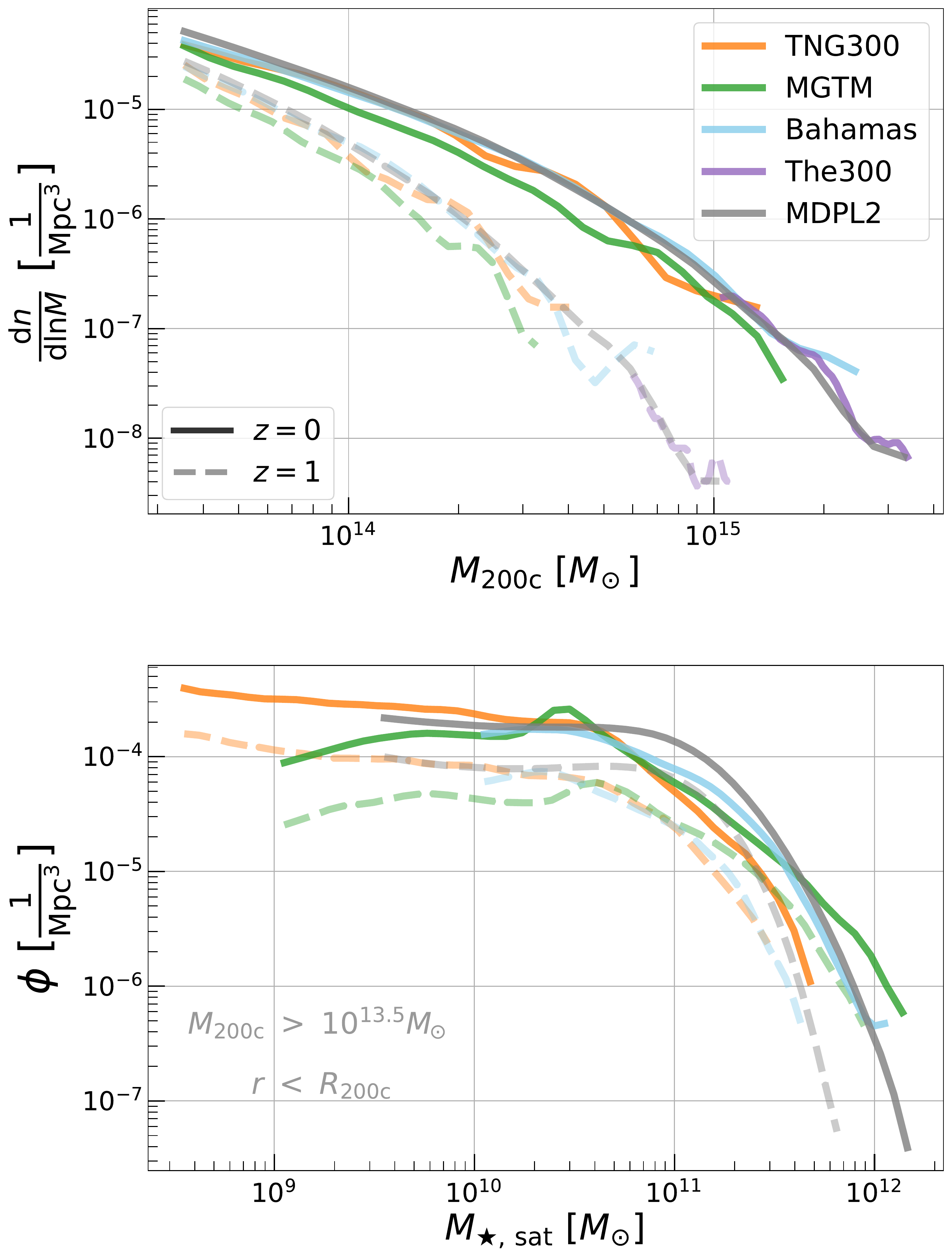}
    \caption{The HMF of host haloes (upper) from the full simulation volume, and the conditional Satellite Galaxy Stellar Mass function (S-GSMF, lower) of all galaxies within $\Rtwohc$ of haloes with $\Mtwohc > 10^{13.5} \msol$, at $z = 0$ (MGTM only is shown at $z = 0.06$). We only show the mass-complete part of The300, and do not include its S-GSMF as it is mass-complete only above $\Mtwohc \gtrsim 10^{15} \msol$ at $z = 0$. We also only show the \textsc{Bahamas} component of BM in both panels, as it is a cosmological halo sample while the complementary \textsc{Macsis} component is not.}
    \label{fig:Mass_functions}
\end{figure}

We use samples of haloes realized in the following five simulations: (i) \textsc{IllustrisTNG}, (ii) \textsc{Magneticum Pathfinder}, (iii) a superset of \textsc{Bahamas} and \textsc{Macsis}, (iv) \textsc{MultiDark Planck 2}, and (v) \textsc{The Three Hundred Project}. Key properties of the simulations are summarized in Table~\ref{tab:simulations}, and a brief description of each follows.

\begin{table*}
	\centering
	\begin{tabular}{l|c|c|c|c|c|c|c|c|c|l}
		Simulation & $L$ [Mpc] & $\Omega_m$ & $H_0\,[\frac{\rm km/s}{\mpc}]$ & $\epsilon_{\rm \,DM}^{z=0} \,\rm [kpc]$ & $m_\star \, [\msol]$ & $m_{\rm DM} \, [\msol]$ & $N_{\rm haloes}$ & $N_{\rm sat,\,tot}$ & Calibration \\
        \hline
            TNG300 & $303$ & $0.3089$ & $67.74$ & $1.48$& $1.1 \times 10^{7}$ & $5.9 \times 10^7$ & $1146$ & $40436$ & See \citet{Pillepich2018Methods} \\
            MGTM & $500$ & $0.2726$ & $70.40$ & $5.33$ & $5.0 \times 10^7$ & $9.8 \times 10^8$  & $4207$ & $90255$ & SMBH, Metals, CL $f_{\rm gas}$ \\
            BM  & $596$ & $0.3175$  & $67.11$ & $5.96$ & $1.2 \times 10^9$ & $6.6 \times 10^9$ & $9430$ & $132334$ & GSMF, CL $f_{\rm gas}$ \\
            MDPL2 & $1475$ & $0.3071$ & $67.77$ & $7.4$ & --- & $2.2 \times 10^9$ & $157051$ & $2339642$ & See \citet{Behroozi2019UniverseMachine} \\
            The300 & --- & $0.3071$ & $67.77$ & $9.6$ & $3.5 \times 10^{8}$ & $1.9 \times 10^9$ & $3180$ & $156662$ & See \citet{Cui2018The300} \\
            \hline
    \end{tabular}
    \caption{Simulation characteristics for the $z = 0$ halo samples. Table adapted from \citet{Anbajagane2020StellarStatistics} with modifications. Note that for MGTM alone we do not have a $z = 0$ dataset and thus characterize the $z = 0.06$ output instead. From left to right, we show: (i) simulation acronym as used in this work, (ii) comoving box size, (iii) cosmic matter density parameter at the present epoch, (iv) Hubble constant, (v) force softening scale, (vi) initial mass of stellar particles, (vii) mass of DM particles, (viii) number of haloes with $\Mtwohc > 10^{13.5} \msol$, (ix) number of satellite galaxies within $r < \Rtwohc$ of host haloes with $\Mtwohc > 10^{13.5} \msol$ and above the minimum stellar mass threshold used for each simulation, and (x) empirical sources used for tuning sub-grid parameters of each simulation, which consist of the Galaxy Stellar Mass Function (GSMF), Supermassive Black Hole scaling (SMBH), Metallicity scaling (Metals), and cluster hot gas mass fraction $<R_{500c}$ (CL $f_{\rm gas}$). All simulations assume a flat $\Lambda {\rm CDM}$ cosmology, with $\Omega_\Lambda = 1- \Omega_m$. See text for references. MDPL2 is an N-body, DMO simulation and does not have stellar particles, and The300 consists of zoom-in (re)simulations of the 324 most massive haloes drawn from MDPL2. 
    }
	\label{tab:simulations}
\end{table*}

The \textsc{IllustrisTNG} project \citep{Springel2018FirstClustering, Pillepich2018FirstGalaxies, Nelson2018FirstBimodality, Naiman2018FirstEuropium, Marinacci2018FirstFields} is a follow up to the \textsc{Illustris} project \citep{Vogelsberger2014Illustris}. It is run with the moving mesh code \texttt{AREPO} \citep{Springel2010EMesh}, and includes a full magneto-hydrodynamics treatment with galaxy formation models, as detailed in \citet{Weinberger2017Methods, Pillepich2018Methods}. We use the highest-resolution run from the TNG300 suite for our main analysis, but also utilize the lower resolution runs in an appendix to perform numerical resolution tests. Haloes are identified via a friends-of-friends (FoF) algorithm, and subhaloes via the \textsc{Subfind} algorithm \citep{Springel2001Subfind, Dolag2009Subfind} which applies a binding energy condition to link particles to substructure. Our galaxy catalogs and host halo properties are obtained/computed from the public data release\footnote{\url{https://www.tng-project.org/}} \citep{Nelson2019TNGPublicData}.

\textsc{Magneticum Pathfinder} \citep[MGTM, ][]{Hirschmann2014MGTM} is a suite of magneto-hydrodynamics simulations run using the smoothed particle hydrodynamics (SPH) solver \texttt{GADGET-3} \citep[last described in][]{Springel2005Gagdet2}. Haloes are identified using a FoF algorithm, and subhaloes are identified using \textsc{Subfind}. We make use of the \texttt{box 2hr} run for this work, and the corresponding galaxy catalogs are obtained from the public database\footnote{\url{http://magneticum.org/data.html}} \citep{Ragagnin2017WebPortal}. Note that we do not have the $z = 0$ catalog for this simulation, and instead use the $z = 0.06$ catalog in its place. Results for other redshifts use the correct catalogs. MGTM also has the most different cosmology to all other simulations in the ensemble (see Table \ref{tab:simulations}); it is based on a WMAP7 cosmology \citep{Komatsu2011WMAP7} whereas all other runs have used Planck cosmologies \citep{Planck2013CosmoParams, Planck2015CosmoParams}.

\textsc{Bahamas} \citep{McCarthy2017BAHAMAS} and its zoom-in companion \textsc{Macsis} \citep{Barnes2017Macsis} --- which we collectively denote with the acronym ``BM'' --- are hydrodynamics simulations run using a version of \texttt{GADGET-3} developed independently of the MGTM version. The \textsc{Macsis} ensemble contains $390$ haloes, with each halo first drawn from a parent $3.2 \gpc$ N-body, dark-matter only (DMO) simulation and then re-simulated in individual, separate volumes with a full hydrodynamics prescription aligned with the \textsc{Bahamas} treatment. \textsc{Macsis} extends the high mass end of BM sample to $\Mtwohc \approx 4 \times 10^{15}\, \msun$ at $z = 0$.
Haloes are once again identified via the FoF algorithm and substructure is identified via \textsc{Subfind}.

\textsc{MultiDark Planck 2} (MDPL2) is part of the MultiDark suite \citep{Klypin2016MultiDark} and is an N-body, DMO simulation run using the \texttt{L-GADGET-2} solver --- a memory-efficient variant of \texttt{GADGET} optimized for simulations with a large number of particles. Haloes and subhaloes were identified using the \textsc{Rockstar} halo finder \citep{Behroozi2013Rockstar}, which identifies structure in the full $6\rm D$ position-velocity phase-space as opposed to the $3\rm D$ position-space used by the other halo/subhalo finders mentioned in this work. While MDPL2 is a DMO simulation with no galaxies (and only subhaloes), galaxies can be ``painted'' onto the subhaloes using the assembly histories of the latter. Galaxy catalogs for MDPL2 have been generated using a variety of different semi-analytical models (SAMs) of which we use public catalogs\footnote{\url{https://www.peterbehroozi.com/data.html}} from the \textsc{UniverseMachine} (UM) prescription \citep{Behroozi2019UniverseMachine}. The host halo quantities are obtained from the public database for MDPL2 \footnote{\url{https://www.cosmosim.org/}} \citep{Riebe2011MultiDarkDatabase}. \textsc{UM}, like many other SAMs, artificially adds so-called ``orphan'' galaxies --- defined as galaxies whose host subhaloes have been tidally destroyed --- back to the catalog so that the two-point galaxy correlation function in the simulated catalog matches observational constraints. 
For these artificially-added galaxies, which reside preferentially near the halo core, \textsc{UM} can only approximately evolve their velocities \citep[see Appendix B]{Behroozi2019UniverseMachine}, and so the velocity statistics of this orphan galaxy population can be discrepant from the ``truth''. 

\textsc{The Three Hundred Project} \citep[The300, ][]{Cui2018The300} is a set of $324$ massive haloes that were first identified in the MDPL2 simulation, and then re-simulated within spheres of radius $22$ (comoving) $\mpc$ with a full hydrodynamics prescription \citep{Rasia2015HydroMethods} using the \texttt{GADGET-X} SPH solver \citep{Beck2016Gadget}. Haloes and subhaloes are identified with Amiga's Halo Finder \citep[\textsc{Ahf},][]{Knollmann2009AHF}, which uses an adaptive mesh refinement grid to represent the density field/contours and also has a binding energy criterion similar to that of \textsc{Subfind}. While The300 is mass-complete only above $\Mtwohc \approx 10^{15} \msol$ at $z = 0$, it still resolves many haloes at masses below this mass scale. We continue using all haloes above $\Mtwohc > 10^{13.5} \msol$ and demonstrate in Appendix \ref{appx:Mass_dependence} that the selection effects in the mass-incomplete part of the halo sample do not impact the quantities of interest to us.

\subsection{Study limitations}

In the strongly non-linear regime of $\Lambda$CDM structure formation, verification studies of the statistics from different simulations is an important way to assess modeling uncertainties. While our ensemble of simulations is extensive, with a variety of  astrophysical treatments and a moderate range in numerical resolution, we note that  in all cases the collisionless dark matter component was evolved using some version of \texttt{GADGET}. \textsc{IllustrisTNG} may seem to be an exception, but its solver, \texttt{AREPO}, inherits its N-body methodology from \texttt{GADGET} as well. Previous results have shown that solutions from different N-body, gravity-only solvers for the matter power spectrum can differ on their estimates of the small scales ($k > 1 \mpc^{-1}$) by more than $3\%$ \citep[see Figure 1]{Schneider2016PowerSpectrumConvergence}. Thus, our ideal simulation ensemble would also include cosmological hydrodynamics simulations based on other N-body solvers, such as \texttt{RAMSES} \citep{Teyssier2002Ramses} and \texttt{PKDGRAV} \citep[last described in][]{Stadel2001PKDGRAV}, but we are not aware of any such simulations that also contain a large enough population of cluster-scale haloes to be used in this work. 

\citet{Mansfield2021SimBias} also show that the scaling relations of internal DM halo properties realized by different suites of N-body simulations do not always converge to the same result --- even those that share the same N-body solver, but differ in their choice of control parameters such as force softening scales and DM particle mass --- with the level of non-convergence varying according to halo property.  We stress that the cluster-scale haloes studied in this work are resolved by at least $N > 10^5$ particles, where \citet{Mansfield2021SimBias} find that velocity-based properties of interest are strongly converged. Further evidence of this comes from the nIFTy comparison project, who simulated the same cluster-scale halo using different codes and found that the bulk halo properties --- such as velocity dispersion and shapes --- agree at the percent-level across codes \citep{Sembolini2016DMOandNR, Sembolini2016NiftyRadiative}. Note that while we also study galaxies resolved by as few as $N \sim 100$ DM particles, we do not focus on the \textit{internal} DM distributions and dynamics of these structures (the primary target for such non-convergence issues) and only concern ourselves with their bulk kinematic properties.

How haloes are identified is also a relevant factor. Throughout our work we rely on catalogs that were constructed by running halo/subhalo finders on a simulation's particle dataset. Differences between the finders can impact the identification of objects in the simulations and affect the derived population statistics. Previous comparison studies of commonly used finders have found some significant differences \citep{Knebe2011HalosgoneMAD, Onions2012SubhalosGoneNotts}. The simulations in our work use a variety of different finders --- three simulations use \textsc{Subfind}, one uses \textsc{Ahf} and one uses \textsc{Rockstar}. Thus, some part of the simulation-to-simulation variance in population statistics will also come from differences across the finders.

%%%%%%%%%%%%%%%%%%%%%%%%%%%%%%%%%%%%%%%%%%%%%%%%%%%%
%%%%%%%%%%%%%%%%%%%%%%%%%%%%%%%%%%%%%%%%%%%%%%%%%%%%%

\section{Methods}\label{sec:Methods}

We employ two different methods to measure the virial (or velocity dispersion) scaling relation for DM and galaxies, respectively. The DM virial scaling (\S\ref{sec:DMVirial}) is obtained by first measuring the DM velocity dispersion for individual haloes and then summarizing the mass-dependent statistics of the population using \textsc{Kllr}, a local linear regression method described further in \S\ref{sec:KLLR}. 
For the galaxy virial scaling (\S\ref{sec:GalaxyVirial}), on the other hand, the sparseness of the satellite galaxy counts motivates us to employ an ensemble likelihood estimator, described further in \S\ref{sec:Galaxy_vel_likelihood}, that circumvents the need to measure the galaxy velocity dispersion for individual haloes. In \S \ref{sec:Validation_Tests}, we sub-sample DM particles and verify that the likelihood method returns scaling parameters consistent with the \textsc{Kllr} estimates.  
The reader wishing to skip the technical details of the measurement can go directly to the definition of velocity bias in \S\ref{sec:vel_bias_model}.

%%%%%%%%%%%%%%%%%%%%%%%%%%%%%%%%%%%%%%%%%%%%%%%%%%%%%

\subsection{Dark Matter Virial Scaling} \label{sec:DMVirial}

By convention \citep{Yahil1977Velocity}, the DM velocity dispersion, $\sigmaDM$, of the host halo is defined as the average of the dispersion along the three Cartesian components,
\begin{equation} \label{eqn:sigma_DM}
    \sigmaDM^2 = \frac{1}{3(N_{\rm part} - 1)}\sum_{i = 1}^{N_{\rm part}} \sum_{j=1}^3 \bigg(v_{ij} - \langle{v}_{j}\rangle \bigg)^2. 
\end{equation}
Here $N_{\rm part}$ is the number of DM particles within $\Rtwohc$ of the halo center, $v_{ij}$ is the velocity of particle $i$ along the $j^{\rm th}$ Cartesian component, and $\langle{v}_{j}\rangle$ is the mean velocity of all $N_{\rm part}$ DM particles along that same component.  All velocities here are peculiar velocities in the proper frame.  

Since the haloes are well-resolved, containing $N \gtrsim 10^{4}$ DM particles, we measure the DM velocity dispersion for each invidiual halo and use a localized linear regression approach, described further below, to infer the $\sigmaDM-\Mtwohc$ scaling relation. 
The study of E08 showed that scaling DM velocity dispersion with $h(z)\Mtwohc$ (instead of $\Mtwohc$) leads to a more universal relation across different cosmologies and redshifts, and so we employ this effective mass as our independent variable. Here $h(z) = H(z)/100$ is the dimensionless Hubble parameter. The utility of this $h(z)$ factor in capturing the cosmological dependence of the DM virial relation was recently confirmed by \citet[see their Section 4.6]{Singh2020CosmologyWithMORs}.

%%%%%%%%%%%%%%%%%%%%%%%%%%
\subsubsection{Kernel-Localized Linear Regression (\textsc{Kllr})} \label{sec:KLLR}

In this work, we employ \textsc{Kllr}\footnote{\url{https://github.com/afarahi/kllr}} \citep[Farahi, Anbajagane \& Evrard, in prep.]{Farahi2018BAHAMAS}, a localized linear regression method, to derive the scale-sensitive estimates of the mean, slope, and intrinsic scatter of the DM virial relation.
The \textsc{Kllr} method is particularly useful for estimating scaling relations in simulations that include baryonic processes. Recently, \citet*{Anbajagane2021BaryonImprints} used \textsc{Kllr} to study the population statistics of key DM halo properties --- including $\sigmaDM$ --- across six decades in halo mass and showed that all relations have clear mass-dependent features that originate from galaxy formation processes.

In brief, \textsc{Kllr} applies a weight to the halo population --- where the weight is given by a Gaussian kernel in log-mass, $\log_{10} M$, with variance $\sigma_{\rm KLLR}^2$ --- and then performs linear regression on the weighted dataset.  Systematically shifting the center of the kernel provides mass-dependent estimates of the fit parameters.  The kernel width is a free parameter and for this work we choose $\sigma_{\rm KLLR} = 0.3 \,\rm dex$. In general, wider kernels wash out small scale features while smaller kernels increase the noise of the estimates. Our choice here is optimized to reduce the uncertainty of the estimates while still capturing the relevant parameter evolution with halo mass.

Given that we use a Gaussian kernel, the \textsc{Kllr} estimates at a mass-scale $M$ are still informed by haloes with $\Mtwohc < M$. So, \textsc{Kllr} parameters at the mass threshold of our analysis, $10^{13.5} \msol$, must be estimated from a sample that extends sufficiently below this value. We therefore include haloes with masses, $\Mtwohc \ge 10^{13} \msol$, which covers the lower $\approx 2\sigma_{\rm KLLR}$ tail of a Gaussian kernel centered on $10^{13.5}\msol$. We have confirmed that this choice leads to a negligible edge-effect bias of $<0.1\%$ in the expectation value of $\sigmaDM$.

%%%%%%%%%%%%%%%%%%%%%%%%%%%
\subsection{Galaxy Virial Scaling} \label{sec:GalaxyVirial}

The quantity available to spectroscopic measurements is the line-of-sight velocity of galaxies within a cluster.  Since the cosmic distances to clusters are typically much larger than their sizes, the radial component of the velocities is close to a simple Cartesian projection.  Measured along a single projection direction, the quantity of interest is the peculiar velocity difference, 
\begin{equation} \label{eqn:delta_gal}
  \Delta v_{ik} = v_i - v_{{\rm halo},k} , 
\end{equation}
of a galaxy $i$ lying within $\Rtwohc$ of the center of host halo $k$. Here, $v_{{\rm halo},k}$ is the host halo $k$'s mean velocity along the chosen projection direction, computed as the mass-weighted mean velocity of all matter components (DM and all phases of baryons) within $\Rtwohc$.\footnote{For MDPL2, which is a DMO simulation, $v_{\rm halo}$ is computed using only DM particles.} Note that this halo reference velocity can differ very slightly from the reference used previously for DM, as the latter is computed using only DM particles whereas here we include the baryonic component as well. However, the characteristic scale of this difference --- estimated using TNG300 --- is small, at $\approx 15 \kms$. 

Again, we do not include the effect of the Hubble flow in the velocities; its inclusion changes the scaling relation parameters by $\approx 0.1\%$ and is thus negligible in our analysis. We use all three orthogonal Cartesian axes to triple the number of independent measurements for each host halo. Thus, the effective halo sample size per simulation in analyses of $\sigmaSat$ is three times that shown in Table~\ref{tab:simulations}.

%%%%%%%%%%%%%%%%%%%%%%%%%%
\subsubsection{Ensemble Velocity Likelihood (EVL)}
\label{sec:Galaxy_vel_likelihood}

Given the sparseness of satellite galaxy counts per halo, particularly at high galaxy stellar mass thresholds, we do not estimate galaxy velocity dispersion for individual haloes like we do for DM.  
This is because the standard deviation is a biased estimator of $\sigmaSat$ when the galaxy count per halo is low, which can be a common occurrence in our analyses. Alternative estimators, such as GAPPER and bi-weight, can provide a more accurate estimate of the dispersion \citep{Beers1990GAPPER}, with the level of accuracy varying across estimators \citep{Ferragamo2020VelDispEstimators}. 
However, one must also make robust estimates of the uncertainty in each measurement in order to properly infer scaling parameters via regression.  This can be a particularly difficult task when the halo is sampled by only two or three galaxies.  

To avoid these sparse-sampling complications, we use an extended version of an ensemble likelihood originally developed by \citet{Rozo2015redmapperIV} to assess cluster membership and then used by F16 and \citet{Farahi2018XXL} for mass estimation.
The basis of the method is an aggregate model for the set of satellite galaxy relative velocities, equation~\eqref{eqn:delta_gal}, conditioned on host halo mass. For fixed host halo mass and redshift, we write an ensemble population likelihood for the collection of 1D galaxy--halo relative velocities. This likelihood is modeled as a convolution of a Gaussian, representing the thermal bath (or velocity distribution) of a single halo, and a log-normal, representing the range of temperatures (or velocity dispersions) at a given halo mass,
\begin{equation} \label{eqn:Vel_likelihood}
    P(\Delta v_{ik}\,|\,M_k,\, \vec{\theta}) = \int P(\Delta v_{ik}\,|\,\sigma_k) \times P(\sigma_k\,|\,M_k,\,\vec{\theta}) \,d\sigma_k,
\end{equation}
where index $k$ runs over all host haloes, and index $i$ runs over the satellite galaxies of host halo $k$. Here, $P(\Delta v_{ik}\,|\,\sigma_k)$ is a Gaussian distribution with zero mean and variance, $(\sigma_k)^2$. The distribution, $P(\sigma_k\,|\,M_k,\,\Vec{\theta})$ is modelled as a Gaussian in $\log_{10} \sigma_k$, with a mean $E[\log_{10}\sigma_k | \, M_k]$ --- described below in equations \eqref{eqn:Scaling_relation} and \eqref{eqn:Scaling_relationQ} --- and a constant, mass-independent variance, $\epsilon^2$.  

The vector, $\Vec{\theta} = \{\pi,\,\alpha,\,\epsilon\}$, contains the log-mean normalization, slope and intrinsic scatter of the pure power-law scaling relation,
\begin{equation}\label{eqn:Scaling_relation}
    E_{\rm lin}[\log_{10}\sigmaSat \,| \, M , \, \Vec{\theta}] = \pi + \alpha\log_{10}\bigg(\frac{h(z)M}{10^{14} \msol} \bigg)
\end{equation}\\
where  $\sigmaSat$ is the satellite galaxy velocity dispersion in $\kms$, $\pi$ is the normalization in decimal log and $\alpha$ is the slope.  We choose $10^{14} \msol$ as the pivot mass scale because it lies close to the midpoint of the halo mass ranges spanned by the different simulations.

We have also extended our model to include a quadratic term with a coefficient, $\beta$, that captures any ``running'' of the power-law slope with host halo mass,
\begin{align}\label{eqn:Scaling_relationQ}
    E_{\rm quad}[\log_{10}\sigmaSat \,|\,  M ,\, \Vec{\theta}] = & \,\,E_{\rm lin}[ ... ]  \nonumber\\
    & + \beta\left[ \log_{10}\bigg(\frac{h(z)M}{10^{14} \msol}\bigg) \right]^2.
\end{align}
However, upon constraining the $\sigmaSat - \Mtwohc$ relation for different halo populations using equation \eqref{eqn:Scaling_relationQ}, we find that $|\beta| < 0.01$ and that the parameter's 68\% confidence intervals always contain $\beta = 0$. Thus, the data shows no preference for the quadratic term and so $\beta$ is not included in the results shown in Section \ref{sec:Galaxy_vel_disp} nor in the parameter set, $\Vec{\theta}$. On the other hand, our validation tests with DM velocity dispersion, presented below, indicate support for $\beta = 0.01 \pm 0.006$ which deviates moderately from $\beta = 0$ at $1.7 \sigma$ significance. 

In principle, one could also add other halo properties to the scaling relation; studies of the \textsc{IllustrisTNG} simulations find that the scatter in $\sigmaDM$ is strongly correlated with secondary DM halo properties such as concentration and shape \citep*{Anbajagane2021BaryonImprints} and so it is plausible, though not necessary, that these secondary properties correlate with $\sigmaSat$ as well. \citet*{Anbajagane2021BaryonImprints} also show that for the halo mass-scales of our study, the intrinsic scatter in $\sigmaDM$ can be reduced by a factor of two when concentration and velocity anisotropy are included in the regression.

Using equations \eqref{eqn:Vel_likelihood} and \eqref{eqn:Scaling_relation}, we can estimate the scaling relation parameters via Bayesian inference. In our model, posteriors for the parameters are obtained after marginalizing/integrating over the distributions of $\sigmaSat$; one distribution per halo as denoted by the integral in equation \eqref{eqn:Vel_likelihood}.
For halo samples of $N = 10^3 - 10^4$, this marginalization step leads to a high-dimensional sampling problem and so we employ the Hamiltonian Monte Carlo method to efficiently sample this space of parameters. Our implementation makes use of existing routines provided by the \textsc{PyMC3} open-source python package\footnote{\url{https://docs.pymc.io/}} \citep{Salvatier2015Pymc3}.

\subsection{Validation of the EVL Estimator} \label{sec:Validation_Tests}

\begin{table}
    \centering
    \begin{tabular}{c|c}
        Name & Method \\
        \hline
        \hline
        Model I & Linear Regression (LR)\\
        Model II & Linear EVL, equation \eqref{eqn:Scaling_relation}\\
        \hline
        Model III & KLLR (Section \ref{sec:KLLR})\\
        Model IV & Quadratic EVL, equation \eqref{eqn:Scaling_relationQ}\\
        \hline
    \end{tabular}
    \caption{The methods corresponding to each model used in the validation tests shown in Figure \ref{fig:Method_comparisons}.}
    \label{tab:Models}
\end{table}
\begin{figure}
    \centering
    \includegraphics[width = \columnwidth]{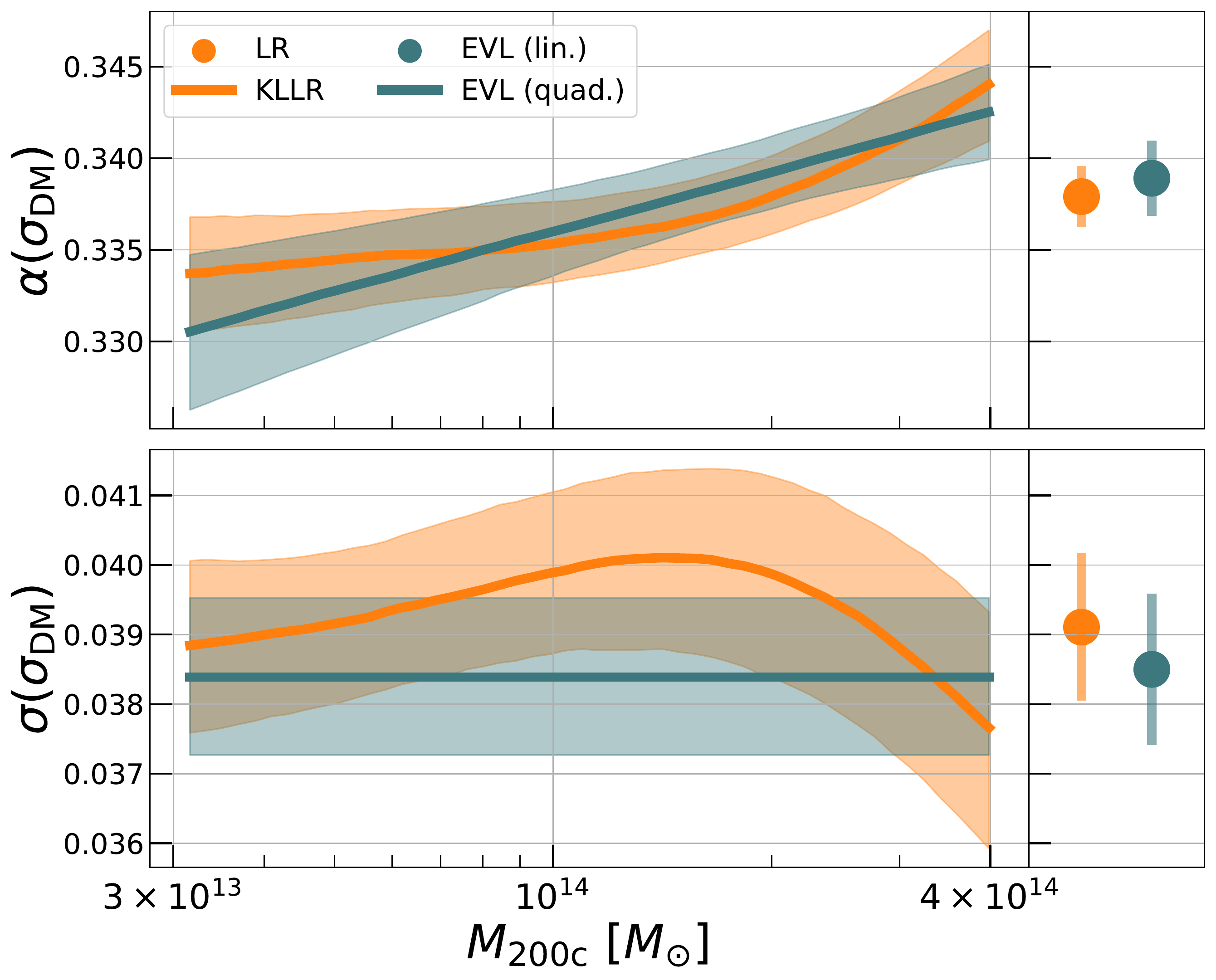}
    \caption{Comparisons of the scaling parameters of DM velocity dispersion with halo mass, extracted via least squares linear regression (LR, Model I), local linear regression (\textsc{Kllr}, Model III), and EVL with a constant slope (lin., Model II) or a running of the slope with log-mass (quad., Model IV). Model I and II, as well as model III and IV, are in statistical consistency for both the slope (top panel) and the scatter (bottom panel). The mean relations of the pairs of models (in Figure \ref{fig:Subsampling}) are also statistically consistent with each other. The rest of this work uses the EVL model with a linear, constant slope. The uncertainties are 68\% intervals determined via bootstrap sampling for models I and III, and the marginalized 1D posteriors for models II and IV.}
    \label{fig:Method_comparisons}
\end{figure}

The mean velocity bias of a halo population depends on the $\sigmaSat - \Mtwohc$ and $\sigmaDM - \Mtwohc$ scaling relations, which are estimated using EVL and \textsc{Kllr}, respectively. Here, we use the DM particles in the TNG300 simulation at $z = 0$ to demonstrate consistency in the scaling parameters returned by the two methods.  For our baseline results, we use both a simple least squares linear regression and \textsc{Kllr}. In this case, $\sigmaDM$ is first computed for individual haloes using all available DM particles in each halo and the regression methods are used to estimate the scaling relation.
The EVL method we are validating is the same as in equation~\eqref{eqn:Vel_likelihood} and equation~\eqref{eqn:Scaling_relation} but the inputs are now DM particle velocities, not satellite galaxy velocities. We randomly select 100 DM particles from each halo and input the velocities from all three Cartesian directions since we are computing the \textit{isotropically-averaged} --- and not line-of-sight --- DM velocity dispersion. Thus each halo is associated with 300 different velocities.

Figure~\ref{fig:Method_comparisons} shows a comparison between the different models, which are listed in Table \ref{tab:Models}. The right panels compare the linear regression and linear EVL (models I and II) which return single values of slope and scatter, and the left ones compare \textsc{Kllr} and the quadratic EVL (models III and IV). The slopes (top panels) and the scatter (bottom panel) are in statistical agreement for both pairs of model comparisons. The normalizations are also statistically consistent, as shown and discussed in Appendix \ref{appx:Subsampling}; this is true even when we input only $N = 10$ DM particles per halo in the EVL method.

To make a fair comparison between these methods, we ensure their estimates are derived from the same sample of haloes, and thus all methods only use haloes with $\Mtwohc > 10^{13.5} \msol$. This notably leads to an edge effect at the lower mass threshold of the \textsc{Kllr} estimates in Figure~\ref{fig:Method_comparisons}, observed as the plateauing of the slope. We reiterate that all fiducial \textsc{Kllr}-based estimates in this work (\eg Figure~\ref{fig:LLR_Summary_Sigma_DM}) are derived from samples with appropriate halo mass ranges and do not suffer from any edge effects.

%%%%%%%%%%%%%%%%%%%%%%%%%%%%%%%%%%%%%%%%%%%%%%%%%%%%%
\subsection{Velocity Bias Definition}\label{sec:vel_bias_model}

The conventional definition of velocity bias is the ratio of the galaxy and DM velocity dispersions conditioned on host halo mass and redshift.  Given the population statistics methods described above, we measure the velocity bias scaling relation, $b_v$, using the difference in the log-linear virial scaling relations, 
\begin{align}\label{eqn:Vel_bias_model}
    \log_{10} b_v = &\,\, E[\log_{10} \sigmaSat \,|\, \Mtwohc, \Mstarsat, z] \nonumber\\  
    & - E[\log_{10} \sigmaDM \,| \, \Mtwohc, z] .
\end{align} 
where $E[x]$ represents the expectation value of $x$ derived by either the \textsc{Kllr} or the EVL method.  
Here, $b_v$ is implicitly a function of $\Mtwohc$, $\Mstarsat$, and $z$.

%%%%%%%%%%%%%%%%%%%%%%%%%%%%%%%%%%%%%%%%%%%%%%%%%%%%%%%%%%%%%%%%%%%%%%%%%%%%%%%%%%%%%%%
%%%%%%%%%%%%%%%%%%%%%%%%%%%%%%%%%%%%%%%%%%%%%%%%%%%%%%%%%%%%%%%%%%%%%%%%%%%%%%%%%%%%%%%

\section{Virial Scaling Relations and Velocity Bias} \label{sec:Results}

We first present results for the DM and galaxy velocity dispersion scaling relations in \S \ref{sec:DM_vel_disp} and \S \ref{sec:Galaxy_vel_disp} before showing the velocity bias scaling relation in \S \ref{sec:Velocity_Bias}. Our scatter is expressed as a natural log --- thus, a fractional scatter --- and uncertainties on all estimates are $68\%$ confidence intervals, determined from the marginalized posteriors for the likelihood-based estimates (for galaxies) or from $1000$ bootstrap resamplings of the \textsc{Kllr} sample (for DM).

\subsection{DM velocity dispersion, \texorpdfstring{$\sigmaDM$}{sigmaDM}} \label{sec:DM_vel_disp}

\begin{figure*}
    \centering
    \includegraphics[width = 2\columnwidth]{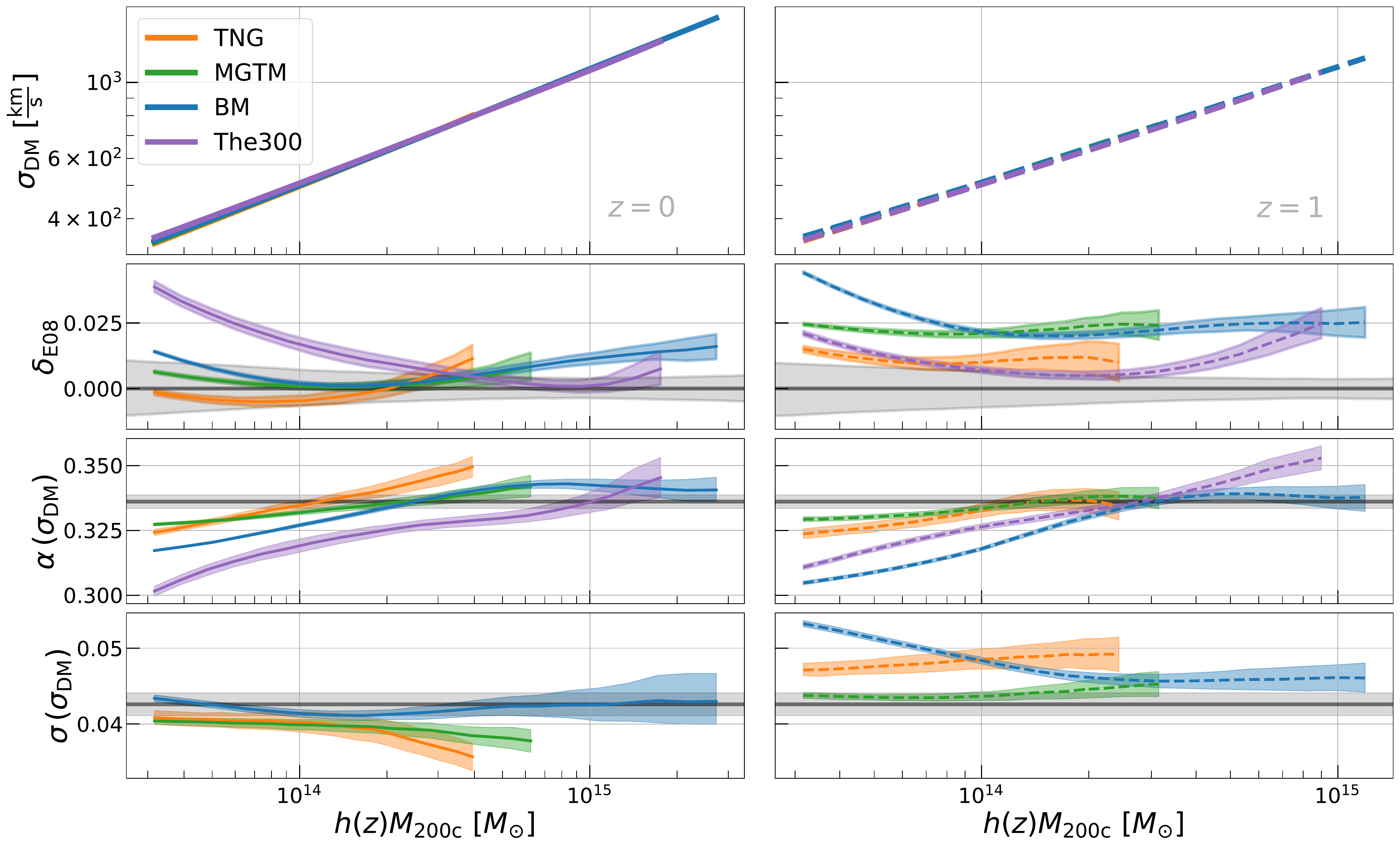}
    \caption{\textsc{Kllr}-derived scaling parameters of the DM velocity dispersion with halo mass at $z = 0$ (left column) and $z = 1$ (right column), from four different simulations (see legend). We follow \citet[E08]{Evrard2008VirialScaling} in using $h(z)\Mtwohc$ as the mass scale; see text for details. The upper middle panel shows the fractional difference between each simulation and the E08 prediction, $\delta_{\rm E08} = \ln(\sigmaDM/\sigmaDM^{\rm E08})$, with colored (gray) bands showing the uncertainty in the simulation (E08) estimates. The slopes (lower middle panel) at low halo masses are shallower than the $\alpha = 1/3$ self-similar expectation, and at high masses agree well with the E08 result, $\alpha = 0.3361 \pm 0.0026$, shown in gray. The scatter (bottom panel) is relatively similar across the simulations and generally consistent with the E08 result, $0.0426 \pm 0.0015$, also shown in gray. \textsc{The300} has a significantly larger scatter at low halo masses 
    due to the environment-selected nature of the low mass sample, and so we omit it from the bottom panel for clarity but provide additional details in Appendix~\ref{appx:Mass_dependence}. 
    }
    \label{fig:LLR_Summary_Sigma_DM}
\end{figure*}

Under self-similar evolution of haloes in virial equlibirum, the slope of the $\sigmaDM - \Mtwohc$ relation is expected to be $\alpha = 1/3$ \citep{Kaiser1986SelfSimilar, BryanNorman1998SelfSimilar}. This expectation has since been confirmed for DMO and non-radiative simulations by \citet{Evrard2008VirialScaling}, whose meta-analysis found $\alpha = 0.3361 \pm 0.0026$.  Simulations with full baryonic physics treatments of galaxy formation find similar results for high mass haloes \citep{Lau2010BaryonDissipationVelDisp, Armitage2018CEagleVelBias}, but deviations of order 10\% in amplitude are found as one moves toward lower mass haloes \citep*{Anbajagane2021BaryonImprints}.  

In Figure~\ref{fig:LLR_Summary_Sigma_DM}, we show the $\sigmaDM - \Mtwohc$ relation of four simulations for $z = 0$ and $z = 1$. MDPL2 is omitted as we do not have access to the requisite data. Note that we regress against $h(z)\Mtwohc$ as, under self-similar evolution, the normalization of $\sigmaDM$ when using this effective mass scale should have no redshift evolution \citep{Kaiser1986SelfSimilar, Evrard2008VirialScaling, Singh2020CosmologyWithMORs}. The normalizations are generally in good statistical agreement with E08 at $z = 0$ (upper middle panel, Figure~\ref{fig:LLR_Summary_Sigma_DM}), although there is moderate tension with the BM simulation at the high mass end.  

The300 haloes at $z=0$ have normalizations of up to $4\%$ higher than E08 for mass scales $\Mtwohc < 10^{14.5} \msol$. The sample at these lower halo masses is incomplete, as they only contain haloes that are within $22 \mpc$ of the 324 most mass haloes from MDPL2 that were selected for resimulation in The300. Thus, these lower mass haloes preferentially lie in regions with strong gravitational tidal fields, and some will have experienced fly-throughs or near encounters with their larger neighbours.

At $z=1$, discrepancies of up to 2\% exist at $\Mtwohc > 10^{14} \msol$, and BM deviates by 4\% at the lowest masses. These deviations are significant at the $>4\sigma$ level in MGTM and BM, but the TNG300 and The300 normalizations exhibit less tension.  We note that E08 quoted a normalization uncertainty of $0.5\%$ at $10^{15} \msol$, rising to 1\% at $10^{13.5} \msol$. Oddly, a normalization upturn at low masses is seen at $z = 1$ in BM but not in The300.

The slopes in the lower middle panel show a clear mass-dependence, as anticipated analytically \citep{Okoli2016Concentration}, with nearly $5\%$ variation across the whole mass range, and with most of the deviations coming at low halo masses.
This justifies our use of the \textsc{Kllr} method over a regular linear regression. The redshift evolution of this feature does have some discrepancies --- the BM runs have shallower slopes at higher redshifts, whereas all other simulation populations have slopes either steeper or comparable to their $z = 0$ slopes.

We note the relevance of multiple full physics simulations exhibiting a mass-dependent slope across the range of redshifts probed in this analysis. The study of \citet*{Anbajagane2021BaryonImprints}, which spans six decades in halo mass across three \textsc{IllustrisTNG} simulation volumes, finds that the population statistics of multiple DM properties --- velocity dispersion, concentration, halo shapes, formation histories etc. --- contain non-monotonic, mass-localized features which originate from the interplay between AGN feedback and gas cooling. For $\sigmaDM$, the inclusion of such processes results in the slope decreasing to $\alpha < 1/3$ toward group-scale haloes, and here we confirm consistent behavior in three other hydrodynamical simulations.

\subsection{Galaxy velocity dispersion, \texorpdfstring{$\sigmaSat$}{sigmaSat}} \label{sec:Galaxy_vel_disp}

\begin{figure*}
    \centering
    \includegraphics[width = 1.8\columnwidth]{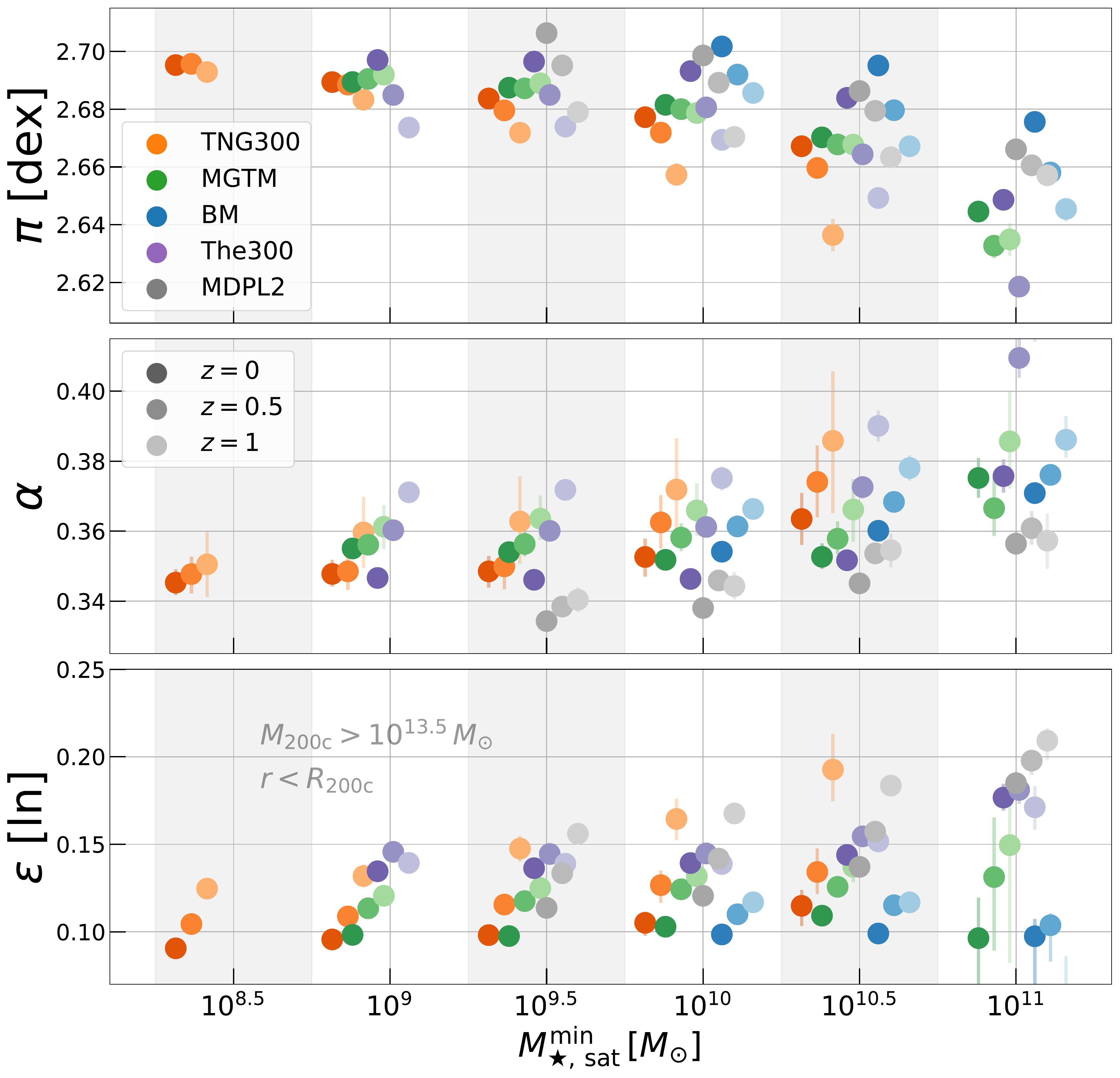}
    \caption{EVL-derived parameters of the $\sigmaSat-\Mtwohc$ scaling relation, equation~\eqref{eqn:Scaling_relation}, as a function of galaxy stellar mass threshold (denoted here as $\Mstarsat^{\rm min}$), for all simulations at redshifts, $z\in \{0, 0.5, 1\}$. We use all galaxies within $\Rtwohc$ of a host halo, and all host haloes with $\Mtwohc > 10^{13.5} \msol$; the exception is \textsc{MDPL2} whose large halo population is subsampled, while preserving the shape of the HMF, to keep our computation time manageable. All simulations display the ``brighter is cooler'' effect; normalizations (top, in decimal log) consistently decrease with increasing $\Mstarsat$ threshold, and slopes (middle) consistently increase. The scatter (bottom, in natural log) shows some dependence on $\Mstarsat$ threshold, particularly at higher redshifts. Most of the errors bars are smaller than the size of the circles. We artificially offset the points horizontally to enhance their visibility. The brackets in the axis labels of the top and bottom panels denote the log base of the quantities shown in each panel (either natural log, $\ln$, or decimal log, $\rm dex$).}
    \label{fig:Param_evolution}
\end{figure*}

In moving to the $\sigmaSat$ relation, we examine the linear parameters for a pure power-law assumption for EVL, and utilize a range of stellar mass thresholds. For TNG300 we do not show results for $\Mstarsat > 10^{11} \msol$ due to the small sample size. Some simulations are unavailable at low $\Mstarsat$ due to resolution limits. 

Figure~\ref{fig:Param_evolution} shows the derived parameters at $z \in \{0,0.5, 1\}$.  All simulations --- both hydrodynamics and semi-analytic variants --- show a clear trend of the scaling relation normalization, $\pi$, decreasing as we increase the stellar mass threshold of the galaxy sample (top panel). This is qualitatively consistent with the signal found in the observational and simulation-based works previously noted (discussed further in Section \ref{sec:Velocity_Bias}), and is therefore also in tension with G15. The drop in normalization also steepens considerably beyond $\Mstarsat > 10^{10.5} \msol$, and this agrees with similar transition points from previous work\footnote{For these comparisons, we have used the TNG300 galaxy catalog to find the absolute magnitudes, in different bands, corresponding to $\Mstarsat = 10^{10.5} \msol$.} --- r-band magnitude $\Mr \approx -21.5$ \citep{Adami1998BrighterToCooler, Adami2000RedshiftEvol}, z-band magnitude $M_z \approx -22$ \citep{Goto2005BrighterToCooler}, and i-band magnitude $M_i \approx -22$ \citep{Old2013VelBias}, where the last result is from simulations whereas the rest are from observations.

The physical origin for the dependence of the normalization on $\Mstarsat$ requires closer investigation, but 
dynamical friction \citep{Chandrasekhar1943DynamicalFriction} has been shown to reduce the velocities of a satellite galaxy sample over time \citep{Ye2017IllustrisVelBias, Armitage2018CEagleVelBias}. 
Satellite galaxies of a larger mass (or mass-proxy, such as $\Mstarsat$) experience stronger dynamical friction, and this naturally leads to the normalization of $\sigmaSat$ decreasing towards more massive galaxies. 
In addition, massive central galaxies are born at rest in their parent haloes, and therefore form a cooler sub-population during mergers compared to their satellite counterparts.  After merging, most of the previously central galaxies will be classified as satellites of the larger system, and the most massive of these satellites would have established a lower velocity due to their prior role as centrals in the pre-merger phase. 

At fixed $\Mstarsat$ and $z$, the simulations' normalizations vary by about $0.02\,\, \rm dex$, or $\sim 5\%$. There is some striation between simulations, with BM preferring a higher normalization than most others, possibly due to its lower resolution. In Appendix \ref{appx:Res_test}, we use multiple TNG300 runs to study resolution effects and show the normalization is amplified by decreased resolution. Such resolution effects also show up in other integral halo properties based on satellite galaxies  --- \citet{Anbajagane2020StellarStatistics} studied the $\Mstarsat$-thresholded satellite galaxy counts of massive haloes in multiple simulations (TNG300, MGTM, and BM) and found that increasing resolution can lead to $50 - 90\%$ more galaxy counts at a given host halo mass.

The normalizations in Figure~\ref{fig:Param_evolution} also show stronger redshift evolution at higher $\Mstarsat$. This is expected as the knee of the S-GSMF evolves more dramatically with redshift compared to the low $\Mstarsat$ end. Thus, by fixing the $\Mstarsat$ threshold at high values and varying redshifts, we sample significantly different parts of the S-GSMF. Note also that the velocity dispersion is always lower at higher redshifts, and this can be understood as follows.
Let us define the ``rank'' of a galaxy as its place in the $\Mstarsat$-ordered list of galaxies at a given redshift; a rank of 1 implies the galaxy is the most massive in the sample. For a fixed $\Mstarsat$ threshold, a low redshift galaxy sample contains more low-rank galaxies (``low'' meaning a larger rank) than the high redshift sample. This, in the context of more massive galaxies being kinematically cooler, naturally implies that the higher redshift sample will have a lower velocity dispersion. Thus, the physical picture of dynamical friction discussed previously can explain the redshift evolution of the normalization as well.  

\begin{figure}
    \centering
    \includegraphics[width = \columnwidth]{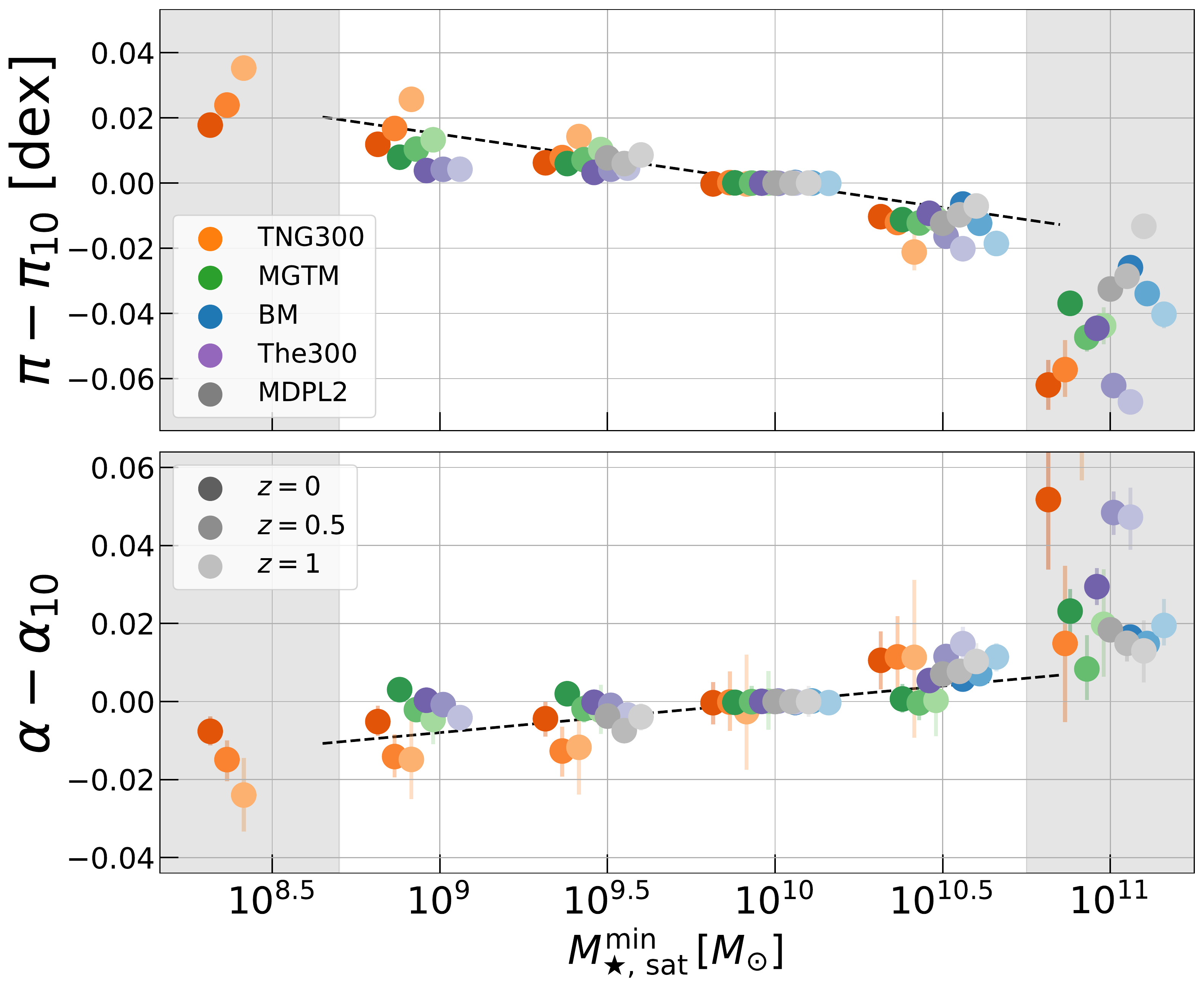}
    \caption{The normalizations (top panel, in decimal log) and slopes (bottom panel) of the $\sigmaSat - \Mtwohc$ scaling relation, as shown in Figure~\ref{fig:Param_evolution}, but offset from their values at $\Mstarsat > 10^{10} \msol$.  All sims show a $5 - 10\%$ difference in both parameters over the three-decades of $\Mstarsat$ threshold values used in this work. The dashed black lines are linear fits, equation~\eqref{eqn:mstar_fit}, with parameters given in Table~\ref{tab:fit_params}. The gray bands show the points that are excluded from the fit.
    }
    \label{fig:Param_evolution_Offsets}
\end{figure}

The slopes of the $\sigmaSat - \Mtwohc$ relation are constrained within $0.33 \lesssim \alpha \lesssim 0.40$ for all simulations across all redshifts and stellar mass thresholds (middle panel, Figure~\ref{fig:Param_evolution}), and have a redshift evolution in agreement with \citet{Munari2013VelDisp}, who found a range $0.35 < \alpha < 0.37$ over $0 < z < 1$ for $\Mstarsat > 10^{9.5} \msol$. \citet{Armitage2018CEagleVelBias} also computed the slopes at different redshifts, but given the large uncertainties on their estimates (resulting from a small sample size) their results are statistically consistent with no redshift evolution. 

Compared to the $\sigmaDM - \Mtwohc$ relation (Figure~\ref{fig:LLR_Summary_Sigma_DM}), 
the $\sigmaSat - \Mtwohc$ relations in all simulations scale more steeply with halo mass. This feature could be due to dynamical friction --- a satellite galaxy of a given size experiences \textit{more} dynamical friction from a less massive host halo \citep[see equation 4.2 and Appendix B]{Lacey1993MergerRates}, and this additional host halo mass-dependence can increase the slope of the $\sigmaSat - \Mtwohc$ relation.  The cold birth persistence after a merger may also play a more significant role in lower mass haloes.  The mismatch of slopes between DM and galaxies naturally results in a halo-mass dependence of the velocity bias (see Section \ref{sec:Velocity_Bias}).

Finally, the scatter depends weakly on $\Mstarsat$ (bottom panel, Figure~\ref{fig:Param_evolution}), though this dependence becomes stronger at higher redshifts. In general, the halo samples at $z = \{0, 0.5\}$ find $\epsilon \approx 0.1$ and this is broadly consistent with previous findings that lie in the range $\epsilon \in [0.06, 0.15]$ depending on galaxy stellar mass and redshift \citep{Munari2013VelDisp, Armitage2018CEagleVelBias}.

\subsubsection{Relative trends with stellar mass threshold}\label{sec:relativeTrends} 

\begin{table}
    \centering
    \begin{tabular}{l|c}
    Parameter   & $q$ \\
        \hline
        Normalization, $\pi$ & $-0.015 \pm 0.001$ \\[2pt]
        Slope, $\alpha$ &  $0.008 \pm 0.001$  \\
        \hline
    \end{tabular}
    \caption{Variation of the scaling relation parameters, $\pi$ and $\alpha$, with stellar mass threshold, equation~\eqref{eqn:mstar_fit}.
    }
    \label{tab:fit_params}
\end{table}

We next focus on the \textit{relative} trends of EVL parameters with $\Mstarsat$. In Figure~\ref{fig:Param_evolution_Offsets}, the normalizations and slopes of each run and redshift, as shown in Figure \ref{fig:Param_evolution}, have been normalized to their values for the threshold $\Mstarsat > 10^{10} \msol$. 
At this threshold, all runs have well-constrained estimates for the parameters. The variation in the normalization and slope with $\Mstarsat$ threshold is well fit by a linear relation,
\begin{equation}\label{eqn:mstar_fit}
    x(\Mstarsat^{\rm min}) - x_{10} \ = \ q \log_{10}\bigg(\frac{\Mstarsat^{\rm min}}{10^{10}\msol}\bigg)
\end{equation}
where $x$ is either $\pi$ or $\alpha$ and $x_{10}$ is the value for the threshold $\Mstarsat > 10^{10} \msol$. 
Fits are performed only using results for $10^{9} \msol < \Mstarsat^{\rm min} < 10^{10.5} \msol$; at lower masses, we only have estimates for TNG300 while at higher masses the data trends deviate significantly from just a simple linear relationship. The data points are also \textit{not} weighted by their errors during fitting as these errors are set by sample size, and so would cause the fitting procedure to strongly weight larger simulations with more haloes while not accounting for numerical resolution.  Instead, in this fit, all simulations are equally weighted regardless of sample size.

The fit parameters for equation \eqref{eqn:mstar_fit} are presented in Table~\ref{tab:fit_params} and the fits are shown in Figure~\ref{fig:Param_evolution_Offsets} as dashed black lines.  The fractional scatter
about the fit is $\approx 1\%$ in the region $10^{9} \msol < \Mstarsat^{\rm min} < 10^{10.5} \msol$. Given a velocity bias (or galaxy velocity dispersion) estimate from a specific sample with a $\Mstarsat$ threshold, these fits can be used to ``translate'' that constraint to different galaxy stellar mass/magnitude thresholds. Note that our chosen analytical model is simplified, and thus approximate, given it does not include (i) redshift evolution of the parameters, and; (i) quadratic log-linear and higher-order terms in the fit, which are particularly relevant towards high $\Mstarsat$.

\subsection{Galaxy Velocity bias, \texorpdfstring{$b_v$}{b\_v}} \label{sec:Velocity_Bias}

\begin{figure*}
    \centering
    \includegraphics[width = 2\columnwidth]{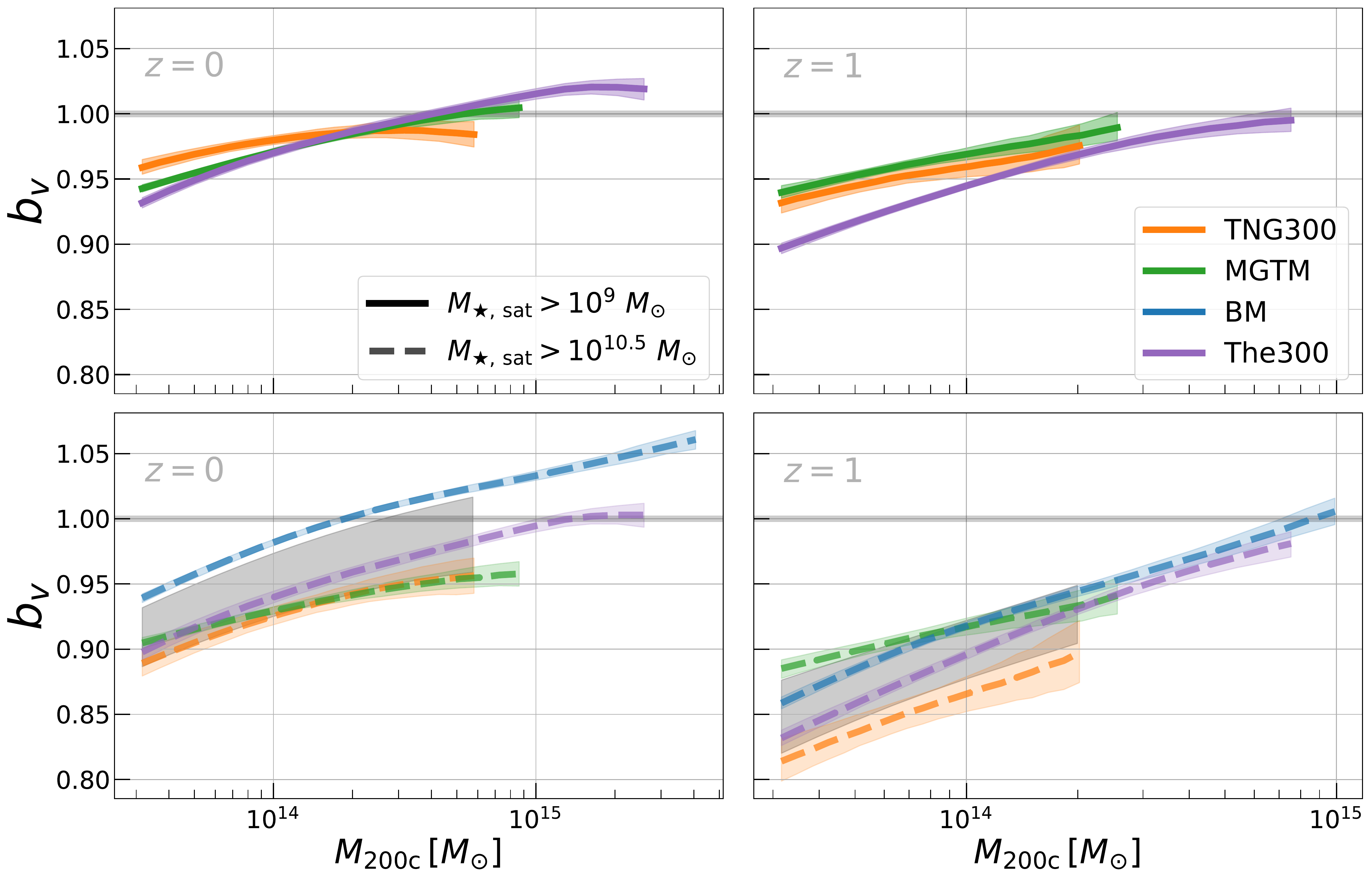}
    \caption{
    The satellite galaxy velocity bias for four simulations (different colors) as a function of halo mass, galaxy stellar mass threshold (different linestyles), and redshift (different panels). The velocity bias, $b_v$, increases with host halo mass, and decreases with redshift and galaxy stellar mass threshold. At fixed $\Mtwohc$, $\Mstarsat$, and $z$, the simulations tend to vary by $\approx 2-3\%$. The horizontal gray lines in all panels shows the unbiased case, $b_v = 1$, and the gray bands in the bottom row show the 68\% confidence interval of our theoretical prior on $b_v$ estimated using all four simulations. 
    }
    \label{fig:Velocity_Bias}
\end{figure*}

The velocity bias, defined in equation \eqref{eqn:Vel_bias_model}, is the ratio of the mean scaling relations of galaxy and DM velocity dispersion with halo mass
(Sections~\ref{sec:Galaxy_vel_disp} and \ref{sec:DM_vel_disp}). Figure~\ref{fig:Velocity_Bias} presents the velocity bias inferred from four simulations, as functions of $\Mtwohc$, for two choices of $\Mstarsat$ threshold, and two choices of redshift, $z$. \textsc{MDPL2} is omitted as we did not have the requisite data for the $\sigmaDM - \Mtwohc$ relation.

At fixed $\Mstarsat$ and $z$, the velocity bias increases nearly linearly with $\log_{10}\Mtwohc$, with $\sim 20\%$ variation over the range presented.  There is good constistency in values among the simulations. At fixed $\Mtwohc$, $\Mstarsat$, and $z$, the variation in $b_v$ across simulations is $2 - 3\%$ for more than 90\% of the 3D parameter space, and improves to nearly \textit{percent-level precision} if we consider the three highest resolution simulations (TNG300, MGTM, and The300). In general, this precision degrades the most at regimes of high $z$, $\Mtwohc$, and/or $\Mstarsat$, where it is amplified by the larger statistical uncertainties due to the smaller sample sizes.

We then construct a theoretical prior for $b_v$ by first representing each simulation's $b_v$ estimate as a Gaussian with a standard deviation given by the statistical uncertainty in $b_v$. Then, we sum the individual Gaussians to form a multimodal distribution, and compute its mean and standard deviation. These provide the moments for a Gaussian representation of the ensemble-based theoretical prior on $b_v$. Examples of these priors are shown in the bottom panels of Figure \ref{fig:Velocity_Bias}.

The stellar mass dependence of $b_v$ comes solely from the $\sigmaSat - \Mtwohc$ relation shown in Section \ref{sec:Galaxy_vel_disp}. Previous observational and simulation-based works have studied the dependence of $\sigmaSat$ (and thus, $b_v$) on different galaxy/subhalo properties and have found similar trends to us. This is because their chosen properties, and subsequent analyses frameworks, are qualitatively related to the $\Mstarsat$ threshold-based analysis we employ here, and we detail these connections below.

Prior observational works all threshold on either absolute magnitudes \citep{Stein1997BrightToCool, Adami1998BrighterToCooler, Adami2000RedshiftEvol, Goto2005BrighterToCooler, Bayliss2017VelDisp, Nascimento2017VelocitySegregation}, or the difference $m - m_3$, where $m$ is the galaxy apparent magnitude and $m_3$ is that of the third brightest galaxy in the cluster \citep{Biviano1992VelocitySegregation, Girardi2003VelocitySegregation, Barsanti2016VelocitySegregation}. Absolute magnitude thresholds are nearly equivalent to $\Mstarsat$ thresholds given the close link between the two quantities, and a threshold on $m - m_3$ is equivalent to a $\Mstarsat$ threshold that is allowed to vary across host haloes. 

Simulation-based works have also used a variety of techniques --- \citet{Biviano2006DynamicalMass} studied $b_v$ separately for early-type and late-type galaxies, and this corresponds to selections of high $\Mstarsat$ and low $\Mstarsat$, respectively.  \citet{Lau2010BaryonDissipationVelDisp, Wu2013VirialScalingPlusBias} select the top $N$ galaxies in each host halo according to $\Mstarsat$ which, like $m - m_3$, is equivalent to using an $\Mstarsat$ threshold that varies across host haloes. \citet{Ye2017IllustrisVelBias, Armitage2018CEagleVelBias} used differential stellar mass bins instead of cumulative ones but this method can still preferentially select galaxies based on $\Mstarsat$. Finally, \citet{Ferragamo2020VelDispEstimators} took all the satellite galaxies belonging to cluster-scale haloes, rank-ordered them according to galaxy mass, and then selected only the top $N\%$. This is equivalent to an $\Mstarsat$ threshold that is set by the galaxy number counts. So all of the above works  --- both observation- and simulation-based --- use frameworks that are qualitatively equivalent to the $\Mstarsat$ thresholds used here, and thus show the same qualitative trends of more massive, or brighter, galaxies being kinematically cooler than their less massive, or fainter, counterparts.

We also find qualitative consistency with previous simulation-based analyses for the trends of velocity bias as a function of host halo mass \citep{Munari2013VelDisp, McCarthy2017BAHAMAS, Ye2017IllustrisVelBias} and redshift \citep{Lau2010BaryonDissipationVelDisp, Munari2013VelDisp}. Note also that the range of values we find, $b_v \in [0.8, 1.1]$, overlaps with those from these previous simulation studies \citep{Lau2010BaryonDissipationVelDisp, Munari2013VelDisp, Wu2013VirialScalingPlusBias, Ye2017IllustrisVelBias, Armitage2018CEagleVelBias, Ferragamo2020VelDispEstimators}.

\section{Revised Mass Scale of Low-\texorpdfstring{$z$}{z} SDSS Clusters} \label{sec:Cluster_Applications}

Uncertainty in cluster mass calibration is a key systematic in cosmological analyses of galaxy cluster abundances \citep[\eg][]{Murata2019HSCMassRichness, Costanzi2021ClusterCosmoDES}. Weak lensing mass calibration is the current gold standard of mass calibration techniques \citep[\eg][]{McClintock2018WLClusterMass, Miyatake2019LensingACT, Murata2019HSCMassRichness, Bellagamba2019KiDsWLMass, Kiiveri2021, Wu2021},
while dynamical mass calibration using ensemble virial scaling is currently limited by the uncertainties in $b_v$ \citep[F16]{Sifon2016VelDispACT}. Increasing the precision of dynamical mass estimation, potentially to the level of weak lensing mass estimation, is also a prerequisite to enabling cluster-based tests of general relativity \citep[e.g.,][]{Shirasaki2021ModGravityClusters}, which require comparisons of weak lensing and dynamical masses. 

In Section \ref{sec:UpdateF16} below, we update the analysis of F16 using our new theoretical prior for $b_v$, and show that we improve the precision on the mean log-halo mass by a factor of $3$. Then, in Section \ref{sec:Roadmap}, we discuss necessary future improvements for further improving the precision of dynamical mass estimates.

We also stress that while we focus on one example (an update to F16) to demonstrate the impacts of our work, improving dynamical mass estimation via our theoretical $b_v$ priors has broader implications for cluster-based science that we do not explore here. For example, \citet[see Section 3 and Section 5.1]{Bocquet2015SPTwithVelDisp} discuss that a $1\%$ prior on the velocity bias improves constraints on both astrophysical and cosmological parameters connected to galaxy clusters by $\approx 30\%$. Cluster counts as a function of their galaxy velocity dispersion and redshift has also emerged as an alternative approach for cluster cosmology \citep{Caldwell2016VelocityCounts, Ntampaka2017TheoryVDF, Ntampaka2019ObsVDF, Kirkpatrick2021SPIDERS}, and the velocity bias is a critical component in forward modelling the relevant observable from cosmological parameters.

\subsection{Updating the Mass -- Richness normalization of F16} \label{sec:UpdateF16}

\begin{figure*}
    \centering
    \includegraphics[width = 2\columnwidth]{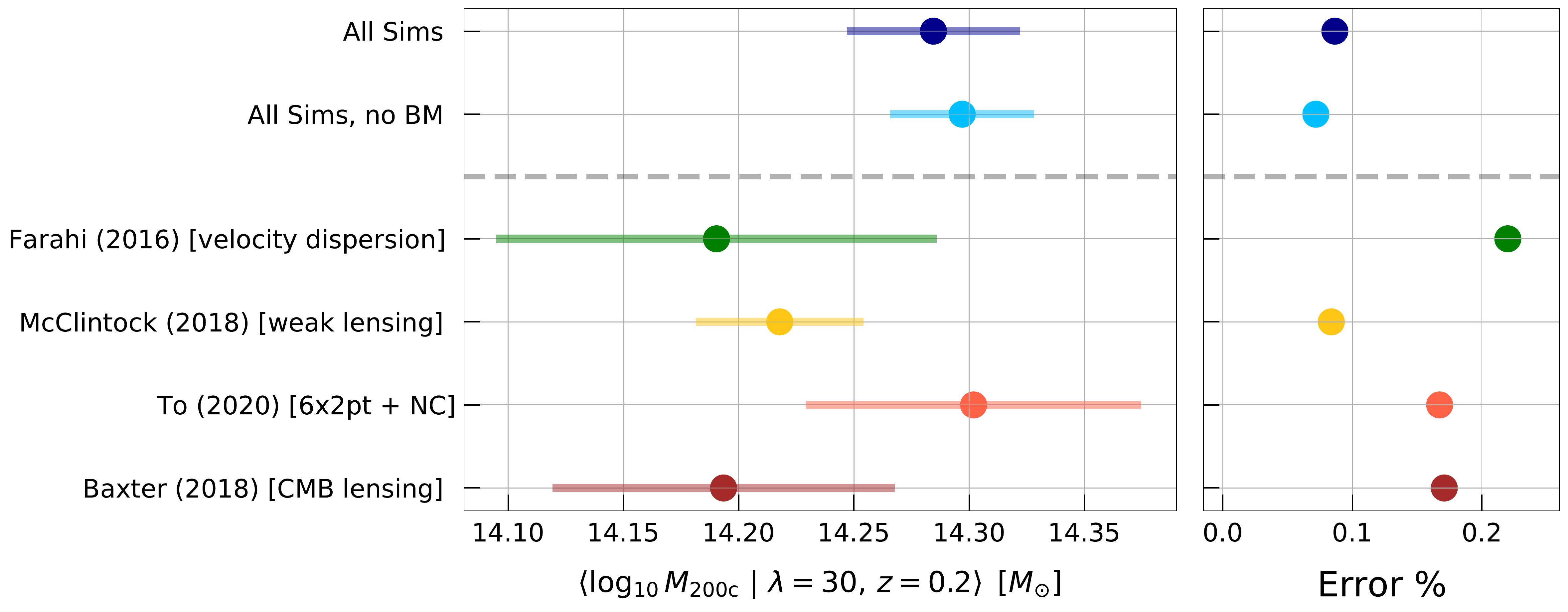}
    \caption{Mass--richness normalization at fixed redshift, $\langle \log_{10}M_{\rm 200c} \,|\,\lambda = 30, z = 0.2\rangle$ derived from different measurement methodologies (left panel, see also Table~\ref{tab:Mass_Constraints}), and the fractional uncertainty on each estimate (right panel).  First two points from the top are our updates to F16 using the theoretical $b_v$ prior from either all four cosmological hydrodynamics simulations or the three highest resolution ones. 
    Either estimate reduces the uncertainty from the original F16 estimate by a factor of $\approx 3$. 
    The last three results were evaluated away from their calibrated pivot points and thus include uncertainties from the evolution with redshift and richness as well.}
    \label{fig:Mass_Constraints}
\end{figure*}

\begin{table*}
    \centering
    \begin{tabular}{l|c|c|c}
        Source & Technique & $\langle \log_{10} M_{\rm 200c} | \lambda = 30, z = 0.2\rangle$ & Error \\
        \hline
        This work, All sims & Ensemble velocity likelihood (EVL) & $14.28 \pm 0.037$ & $8.7\% $\\
        This work, Sims w/o BM & EVL & $14.30 \pm 0.031$ & $7.4\%$\\
        \hline
        \citet{Farahi2016StackedSpectro} & Older EVL & $14.19 \pm 0.096$ & $22\%$\\
        \citet{McClintock2018WLClusterMass} & Background galaxy weak lensing & $14.22 \pm 0.035$ & $8.1\%$\\
        \citet{To2021_6x2+NC} & Galaxy/DM clustering + cluster abundance & $14.30 \pm 0.079$ & $18\%$\\
        \citet{Baxter2018CMBLensingClusterMass} & CMB lensing & $14.19 \pm 0.074$ & $17\%$\\
        \hline
    \end{tabular}
    \caption{The mass-richness normalization from this work and from previous works that estimated the same using different techniques: a slightly older version of EVL (F16), weak lensing \citep{McClintock2018WLClusterMass}, CMB lensing \citep{Baxter2018CMBLensingClusterMass}, and from a combined cosmological analysis of different probes, including cluster-scale haloes \citep{To2021_6x2+NC}. The estimates for \citet{McClintock2018WLClusterMass, To2021_6x2+NC} are quoted away their respective pivot points, and thus their errors include additional uncertainties coming from the redshift and richness evolution. The uncertainties at their pivot scales are $5.1\%$ and $4.6\%$ respectively.}
    \label{tab:Mass_Constraints}
\end{table*}

The work of F16 uses a slightly older version of EVL to estimate the the velocity dispersion of galaxies for an SDSS redMaPPer cluster sample, and then employs the velocity bias constraints from G15 to estimate the normalization of the halo mass--optical richness scaling relation $\langle M_{\rm 200c} \,|\, \lambda, z\rangle$, where the richness $\lambda$ is an observational analog for the counts of red-sequence satellite galaxies in a halo. We will henceforth refer to masses estimated using this approach as ``EVL masses.'' Our update here replaces the G15 estimates with the theoretical $b_v$ prior estimated in this work and recomputes the F16 normalization. While we focus primarily on updating the normalization, the slope of the scaling relation will also shift as G15 (and thus, F16) assumed that $b_v$ did not vary with halo mass, whereas Figure \ref{fig:Velocity_Bias} shows that there is a significant mass-dependence.

The pivot scales of F16 are $z=0.2$ and $\lambda = 30$ for which the normalization, derived with the G15 estimate of $b_v = 1.05$, is $\Mtwohc = 1.56 \pm 0.35 \times 10^{14} \msol$. We are fortunate that we have $b_v$ estimates near this halo mass scale from all four simulations (see Figure~\ref{fig:Velocity_Bias}). The ensemble-estimated theoretical prior for $b_v$ is most accurate in halo mass ranges where all simulations are available, and gets more limited towards high halo masses, with only two simulations (BM and The300) sampling $\Mtwohc > 10^{15} \msol$ at $z = 0$ and $\Mtwohc > 10^{14.5} \msol$ at $z = 1$.

To update the F16 mass normalization, we use the Gaussian representation of the theoretical $b_v$ prior as described in Section \ref{sec:Velocity_Bias} and shown in Figure~\ref{fig:Velocity_Bias}. The first two moments of the Gaussian prior are interpolated over $\Mstarsat$, $\Mtwohc$, and $z$ as necessary. F16 employed an approximate magnitude threshold of $\Mr \sim -21.5$ for their redMaPPer sample, and this corresponds to a stellar mass of $\Mstarsat \approx 10^{10.3} \msol$ in the TNG300 simulation. We tested the inclusion of a 50\% uncertainty ($\approx 0.2 \rm \, dex$) on this threshold, and note that our results are not that sensitive to this choice.  At a halo mass of $10^{14.2} \msol$ at $z=0$, the dependence of velocity bias on stellar mass threshold is weak, $b_v \propto (\Mstarsat)^{-0.015}$. Since the EVL mass estimate scales as $M \propto b_v^{-3}$, a 50\% uncertainty in $\Mstarsat$ translates to an $2.3\%$ error in the mass scale. This is currently not a significant source of uncertainty and alters our total uncertainty estimates, presented further below, by $\approx 0.2\%$.

The velocity bias is also a function of halo mass. 
Given the stellar mass threshold of $\Mstarsat > 10^{10.3} \msol$ at $z=0.2$, we solve for the mass normalization iteratively.  We first make an initial guess for the velocity bias, and derive a mass normalization constraint. Then we compute the simulation ensemble-estimated velocity bias at that mass scale and re-derive the mass normalization. This step is repeated until the normalization converges to within $0.01\%$, which takes $<10$ iterations. 

The mean and uncertainty of the normalization are determined via Monte Carlo sampling. We first draw a large number of random samples --- we choose $N = 10^7$ for which our uncertainty estimates converge to within absolute deviations of $10^{-4}$ --- from Gaussian priors for each of the following: the $\sigmaSat$ measurement from F16, the central galaxy velocity bias used in F16 (which comes from G15), the satellite galaxy velocity bias as constrained in this work, the normalization and slope of the $\sigmaDM - \Mtwohc$ relation from E08, and finally, the cosmological parameters $\Omega_M$ and $h$ from \citet{Planck2015CosmoParams}. The mean and variance we use for each Gaussian prior come from the works listed above alongside each quantity. For $b_v$, these are obtained from the ensemble-estimated theoretical prior. Then, we once again compute the mass normalization iteratively, and the mean and standard deviation of the resulting distribution of $10^7$ normalizations is the quoted mean and uncertainty.

Figure~\ref{fig:Mass_Constraints} presents two variants of the updated F16 constraints  --- one where we use all four simulations (TNG300, BM, MGTM, and The300), and one where we exclude BM and use only the three highest resolution ones.  The latter is motivated by the resolution considerations mentioned previously and discussed further in Appendix \ref{appx:Res_test}. Since the mean velocity bias has declined, to $b_v = 0.98$ compared to the original G15 estimate of $b_v = 1.05$, the inferred mass scale (which goes as $M \propto b_v^{-3}$) rises by nearly 0.1 dex.  The $0.07$ shift in $b_v$ is within the stated $1\sigma$ error of G15 ($\Delta b_v = 0.08$) so the revised mass estimate remains consistent with the wide 68\% confidence interval of the original F16 estimate determined using the G15 constraints. The revised EVL normalization also remains consistent with existing weak lensing estimates, including the recent multi-probe estimate of \citet{To2021_6x2+NC}. 

F16 report a slope of $\alpha = 1.31 \pm 0.06_{\rm stat} \pm 0.13_{\rm sys}$ for the mass--richness relation, but this value was derived assuming a constant $b_v$.  At low redshift, the simulations display a weak halo mass dependence, $b_v \propto (\mtwoh)^{0.03}$, which would imply a shift of $-0.09$, meaning a revised slope of $\alpha \approx 1.22$.  The shallower slope of the SDSS redMaPPer mass--richness relation still lies between values in the literature derived from different methodologies; \citet{Simet2018SDSSMlens} find $\alpha = 1.30 \pm 0.10$ while \citet{Murata2019HSCMassRichness} find $\alpha = 1.00 \pm 0.05$.  

Figure~\ref{fig:Mass_Constraints} and  Table~\ref{tab:Mass_Constraints} compare our revised EVL mass normalization with those from previous observational studies.  
When  a published value is quoted at a different richness and redshift, we translate it to the pivot richness and redshift of F16 --- $\lambda = 30$ and $z = 0.2$ --- while incorporating the extra uncertainty from moving off the fiducial pivot scale due to slope or redshift evolution uncertainties.  When other works only quote the richness--mass normalization, $\langle \lambda \,|\, \Mtwohc, z \rangle$, we use the formalism of \citet{Evrard2014MultiPropertyStatistics} to invert the scaling relation and obtain the mass--richness normalization, $\langle \Mtwohc \,|\, \lambda, z \rangle$. Many studies also quote their halo mass in $\Mtwohm$, which is defined similar to $\Mtwohc$ but now with $\rho_\Delta = 200\rho_m(z)$, where $\rho_m$ is the mean matter density at redshift $z$. We convert $\Mtwohm \rightarrow \Mtwohc$ using the \textsc{COLOSSUS}\footnote{\url{https://bdiemer.bitbucket.io/colossus/}} open-source python package \citep{Diemer2018COLOSSUS} while employing the concentration-mass relation from \citet{Diemer2019concentrations}. The uncertainty from the concentration relation is not incorporated into our final estimate.

The right panel of Figure~\ref{fig:Mass_Constraints} shows the magnitude of the total uncertainty in the various mass normalizations.  The specific values we constrain, as well as those of the comparison works, are found in Table~\ref{tab:Mass_Constraints}. The improved precision on $b_v$, due to the simulation ensemble-estimated theoretical prior, reduces the original F16 uncertainties by a factor of 3, and makes dynamical mass estimation a competitive technique in determining the normalization of the mass--richness relation.

\subsection{Roadmap to more precise EVL mass estimates} \label{sec:Roadmap}

\begin{figure}
    \centering
    \includegraphics[width=\columnwidth]{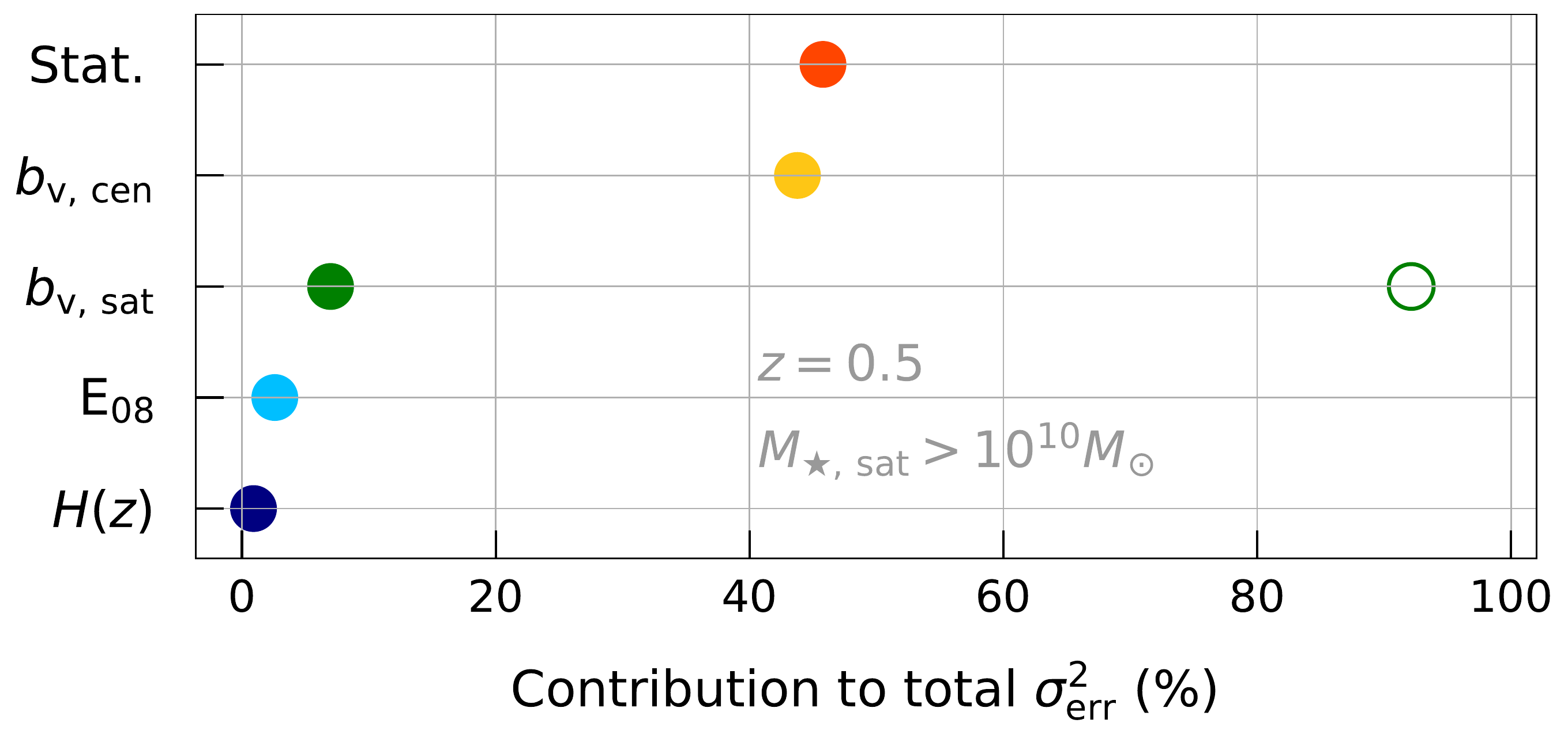}
    \caption{The percentage contribution of each component $X$ to the total \textit{variance} of the mass normalization. From top to bottom $X$ is: (i) the statistical uncertainty from the $\sigmaSat$ observational measurement, (ii - iii) the central and satellites galaxy velocity biases, (iv) the $\sigmaDM - \Mtwohc$ relation from E08, and (v) the cosmology uncertainty from $H(z)$ assuming Planck-like constraints. The open green circle is the previous result using G15. We show results for $z = 0.5$ and $\Mstarsat > 10^{10} \msol$, but the qualitative behavior is similar across a large part of the parameter space. We have used only the three highest resolution simulations (TNG300, MGTM, and The300) for this analysis. See text for details.
    }
    \label{fig:Uncertainty_Contribution}
\end{figure}

To motivate potential improvements in future analysis, we illustrate in Figure~\ref{fig:Uncertainty_Contribution} the importance of difference sources of the uncertainty in the EVL mass normalization.  The figure shows the fraction of the overall variance contributed by the uncertainty in each individual source.  For the purpose of illustration, we assume $z = 0.5$, $\Mstarsat > 10^{10} \msol$, and exclude BM when determining the theoretical prior for $b_{v, \rm sat}$. The statistical uncertainty is the SDSS value from F16.  The overall fractional uncertainty in the mass scale is $\sigma_{\rm err} = 7.8\%$, not including the $\Mstarsat$ uncertainties discussed previously.

The halo velocity we employ as the theoretical reference rest frame in equation~\eqref{eqn:delta_gal} is unavailable to observers.  Instead, F16 uses the central galaxy velocity as a proxy, thereby bringing the central galaxy velocity bias, $b_{v,\rm cen}$, into the analysis.  Uncertainty in that component is comparable to the current SDSS statistical error in the velocity dispersion normalization at $\lambda = 30$. So satellite galaxy velocity bias is now a sub-dominant source of systematic uncertainty, while the central galaxy velocity bias becomes the dominant source.  

Prospects for improvements to the dominant sources of uncertainty are good. The statistical uncertainty of the measurement will improve just by increasing the size and depth of spectroscopic samples.  The original analysis of velocity dispersion scaling with optical richness by \citet{Rozo2015redmapperIV} employed roughly 9000 clusters, each sampled by 20 or more spectroscopic galaxy members, for a sample size of approximately 200,000 galaxies.  
Recent wide-area imaging surveys, such as the Dark Energy Survey \citep[DES,][]{DES2005} and Hyper Suprime-Cam \citep{Aihara2018HSC}, are producing much larger optically-selected cluster samples, and  overlapping areas of sky are being probed by Sunyaev-Zel'dovich observations from the Atacama Cosmology Telescope \citep{Choi2020ACT} and South Pole Telescope \citep{Carlstrom2011SPT}, and also X-ray observations from the eROSITA mission \citep{Merloni2012eROSITA}.  
Spectroscopic surveys such as the Dark Energy Spectroscopic Instrument \citep[DESI,][]{Dey2019OverviewDESI}, Euclid \citep{Laureijs2011EUCLID} and, in the longer term, the Nancy Grace Roman Telescope \citep{Akeson2019Roman} and Extremely Large Telescope MOSAIC \citep{Evans2015ELTMosaic}, will produce samples larger by an order of magnitude or more compared to the SDSS analysis of F16. 

The central galaxy velocity bias, on the other hand, will need to be studied more extensively via simulations to quantify the theoretical uncertainty by constructing an ensemble-estimated theoretical prior in a manner similar to that of this work. Previous works have calibrated this bias as a function of galaxy and/or host halo properties, but either do not adequately describe galaxy velocities within clusters, i.e. the one-halo term (G15), or are unable to quantify the theoretical uncertainty due to the study being limited to a single simulation \citep{Martel2014, Ye2017IllustrisVelBias}.  

Other sources of systematic uncertainty, such as miscentering and projection, will also need to be precisely calibrated.  A promising approach for the former may be to use multi-wavelength observations, such as X-ray and optical \citep{Zhang2019MiscenteringDES}, to define a well-centered subset of clusters.  Application of EVL and other dynamical mass techniques to this subset would produce estimates more reflective of the underlying massive halo population.  

Note that the EVL method focuses solely on a line-of-sight velocity dispersion. Other methods, such as those using caustics \citep{Rines2003Caustic, Rines2006CausticSDSS, Gifford2013Caustics, Gifford2017CausticsStack}, or the Jeans equation \citep{Mamon2013Mamposst}, also employ the transverse radial distances. Recent deep learning techniques also make full use of the 2D phase space consisting of line-of-sight velocities and transverse radial distances \citep{Ho2019VelocityCNN}.

Additionally, while we have discussed and documented the ``brighter is cooler'' effect in the context of a velocity dispersion/bias, the effect impacts the full 6D position-velocity phase space of the galaxies. So other relevant features in this phase space --- such as the outer caustic surface, or splashback feature \citep{Diemer2014Splashback, Adhikari2014Splashback, More2015Splashback} --- will also be impacted by this effect. Previous studies of both observations and simulations have found that the radial location of the splashback feature, as estimated via the galaxy number density profile, depends on galaxy properties such as color and mass \citep{Adhikari2020SplashbackACTxDES, Dacunha2021SplashbackTNG}.

\section{Conclusions} \label{sec:conclusions}

Estimating the mass scale of galaxy clusters from the ensemble velocity statistics of satellite galaxies is a method that is currently limited by uncertainties in how well galaxies trace the DM velocity field.
The velocity bias, $b_v$ --- which is the ratio of velocity dispersion of satellite galaxies to that of dark matter --- is a key source of uncertainty that we address using new statistical methods applied to an ensemble of cosmological hydrodynamics simulations that include an extensive range of galaxy formation physics.  

We extract estimates of $b_v$ as a function of host halo mass, satellite galaxy stellar mass threshold, and redshift using a set of four independent cosmological hydrodynamics simulations. This is done using both a local linear regression, as well as a new ensemble velocity likelihood method that is unbiased for low galaxy counts per halo. The collective analysis of the multiple simulations allows us to derive an ensemble-estimated theoretical prior on $b_v$ that quantifies the uncertainty driven by different astrophysical and numerical treatments. Our main results are as follows:

\begin{itemize}
    
    \item  At $z = 0$, the DM velocity dispersion scaling relation is consistent across all simulations at the one percent level and agrees with prior expectations from E08 (Figure~\ref{fig:LLR_Summary_Sigma_DM}), but larger deviations are seen at $z=1$. The slopes in all simulations have a consistent mass-dependence, and are shallower than the self-similar expectation ($\alpha = 1/3$) at halo masses below $\Mtwohc < 10^{14}\msun$.
 
    \item The normalization of the galaxy velocity dispersion scaling relation decreases with stellar mass threshold, indicating that more massive galaxies are kinematically cooler than their lighter counterparts (top panel, Figure~\ref{fig:Param_evolution}).  The redshift and stellar mass dependence of this feature is consistent across an ensemble consisting of four hydrodynamics simulations and one N-body/semi-analytic simulation (Figure \ref{fig:Param_evolution_Offsets}).
    
    \item In all simulations, the slopes of the $\sigmaSat - \Mtwohc$ scaling relation are greater than the self-similar expectation, and the relation steepens with both stellar mass threshold and redshift (middle panel, Figure~\ref{fig:Param_evolution}).
    
    \item The ratio of the $\sigmaSat - \Mtwohc$ and $\sigmaDM - \Mtwohc$ scaling relations yields a velocity bias, $b_v$, that varies as a function of host halo mass, galaxy stellar mass threshold, and redshift (Figure~\ref{fig:Velocity_Bias}). The simulation-to-simulation variation is $2-3\%$ for more than 90\% of the 3D parameter space constituting $\Mtwohc$, $\Mstarsat$, and $z$. However, this reduces to \textit{percent-level precision} when considering only the three highest resolution simulations (TNG300, MGTM, and The300). The uncertainty is larger at higher redshift and higher halo/stellar mass scales where the halo samples are sparse.
    
    \item We update the mass normalization of optically selected SDSS clusters studied in F16 by using the ensemble-estimated theoretical $b_v$ prior derived in our work. Our more precise estimate improves the uncertainty on the normalization from $22\%$ to $7-8\%$ (Figure \ref{fig:Mass_Constraints} and \ref{fig:Uncertainty_Contribution}), and makes dynamical mass estimation using the ensemble velocity of satellite galaxies a technique that is competitive with weak lensing.
\end{itemize}

The \textit{trends} in velocity bias discussed in this work are all empirically testable with ongoing spectroscopic campaigns of clusters such as SPIDERS \citep{Kirkpatrick2021SPIDERS} and DESI. The dependence of $\sigmaSat$ on $\Mstarsat$ (or more precisely, the galaxy luminosity) has already been observationally studied for many different modestly-sized samples of clusters ($N \sim 100$) as was noted before, while such observational studies of the redshift and halo mass trends have not yet been well-explored. The same datasets could be used to derive EVL-based constraints on the mass--richness normalization for cluster samples selected by different methods. Comparisons of precise mass-scale estimates between X-ray, SZ and optically selected samples would offer insights into sample selection models, the strength of projection effects, and intrinsic covariance among stellar, hot gas and dark matter properties.   

Finally, while we have focused on galaxy cluster mass calibration as the premier application of our velocity bias constraints, our results can also be relevant for models of small-scale RSDs measurements \citep[\eg][]{Tinker2007RSDModelingHOD, Reid2014VelBiasModGravRSD, Guo2015VelocityBiasSDSS, Guo2015RSD3ptFunctions, Guo2015RsdLuminosityDependance, DESI2016, Yuan2018RSD3ptFunctions, Zhai2019RSDEmulator, Tonegawa2020SmallScalesRSD, Alam2021GAMASmallScaleRSD, Lange2021BossLowRedshiftRSD, Shirasaki2021SAMforRSD, DeRose2021RSDwithSHAM} and more generally, any small-scale N-point auto- or cross-correlation function that uses galaxy kinematics as an observational tracer of the DM velocity field. Such cosmological probes will also be highly relevant over the next decade given the expected large sky coverage and redshift range of DESI and future spectroscopic surveys\footnote{\href{https://www.noao.edu/2020Decadal/files/BoltonAdamS2b_SpectroscopicRoadmap.pdf}{\emph{e.g.}, ASTRO2020 White Paper: Towards a Spectroscopic Survey Roadmap for the 2020s and Beyond}}.

\section*{Acknowledgements}

We thank Chun-Hao To and Alexander Knebe for useful discussions. We also thank the \textsc{IllustrisTNG} Team for publicly releasing all TNG simulation data, and Peter Behroozi for doing the same for the \textsc{MDPL2} UniverseMachine catalogs. We also thank the anonymous referee for helpful comments on the presentation of the methods employed in this work.

DA is supported by the National Science Foundation Graduate Research Fellowship under Grant No. DGE 1746045. AF is partially supported by a Michigan Institute for Data Science (MIDAS) Fellowship. WC is supported by the European Research Council under grant number 670193 and by the STFC AGP Grant ST/V000594/1. He further acknowledges the science research grants from the China Manned Space Project with NO. CMS-CSST-2021-A01 and CMS-CSST-2021-B01. KD acknowledges support by the Deutsche Forschungsgemeinschaft (DFG, German Research Foundation) under Germany's Excellence Strategy - EXC-2094 - 390783311 and through the COMPLEX project from the European Research Council (ERC) under the European Union’s Horizon 2020 research and innovation program grant agreement ERC-2019-AdG 860744. GY acknowledges financial support from the MICIU/FEDER (Spain) under project grant PGC2018-094975-C21.

This research was supported in part through computational resources and services provided by Advanced Research Computing (ARC), a division of Information and Technology Services (ITS) at the University of Michigan, Ann Arbor. 

The TNG simulations were run with compute time granted by the Gauss Centre for Supercomputing (GCS) under Large-Scale Projects GCS-ILLU and GCS-DWAR on the GCS share of the supercomputer Hazel Hen at the High Performance Computing Center Stuttgart (HLRS). 

The {\it Magneticum} simulations were performed at the Leibniz-Rechenzentrum with CPU time assigned to the Project `pr86re'.

The 300 project has received financial support from the European Union’s
Horizon 2020 Research and Innovation programme under the
Marie Sklodowskaw-Curie grant agreement number 734374, i.e.
the LACEGAL project.  
We would like to thank The Red Espa{\~n}ola de Supercomputaci{\'o}n for granting us computing time at the MareNostrum Supercomputer of the BSC-CNS where most of the 300 cluster simulations
have been performed. The MDPL2 simulation has been performed
at LRZ Munich within the project pr87yi. The CosmoSim database
(https://www.cosmosim.org) is a service by the Leibniz Institute for
Astrophysics Potsdam (AIP). Part of the computations with \texttt{GADGET-X} have also been performed at the ‘Leibniz-Rechenzentrum’ with
CPU time assigned to the Project ‘pr83li’.

The MultiDark Database used in this paper and the web application providing online access to it were constructed as part of the activities of the German Astrophysical Virtual Observatory as result of a collaboration between the Leibniz-Institute for Astrophysics Potsdam (AIP) and the Spanish MultiDark Consolider Project CSD2009-00064.

All analysis in this work was enabled greatly by the following software: \textsc{Pandas} \citep{Mckinney2011pandas}, \textsc{NumPy} \citep{vanderWalt2011Numpy}, \textsc{SciPy} \citep{Virtanen2020Scipy}, and \textsc{Matplotlib} \citep{Hunter2007Matplotlib}. We have also used
the Astrophysics Data Service (\href{https://ui.adsabs.harvard.edu/}{ADS}) and \href{https://arxiv.org/}{\texttt{arXiv}} preprint repository extensively during this project and the writing of the paper.

\section*{Data Availability}

The tables containing the scaling relations for galaxy/DM velocity dispersions and the velocity bias are publicly available at \url{https://github.com/DhayaaAnbajagane/VelocityBias}. We also provide a convenience script that parses the scaling parameter files, and also provides the theoretical $b_v$ prior while being able to interpolate over host halo mass, galaxy stellar mass threshold, and redshift, as needed.

The galaxy and halo catalogs for \textsc{IllustrisTNG}, \textsc{Magneticum Pathfinder}\footnote{The $\sigmaDM$ and $\Mtwohc$ quantities for \textsc{Magneticum Pathfinder} are not available at the public repository and were provided by one of us for use in this study.} and \textsc{UniverseMachine} are all publicly available at the repositories linked in \S \ref{sec:Data}. The data for \textsc{The300}, \textsc{Bahamas}, and \textsc{Macsis} are not available at a public repository, but can be provided on request.

%%%%%%%%%%%%%%%%%%%%%%%%%%%%%%%%%%%%%%%%%%%%%%%%%%


\bibliographystyle{mnras}
\bibliography{References} % if your bibtex file is called example.bib

%%%%%%%%%%%%%%%%%%%%%%%%%%%%%%%%%%%%%%%%%%%%%%%%%%

%%%%%%%%%%%%%%%%% APPENDICES %%%%%%%%%%%%%%%%%%%%%

\appendix

\section{Resolution effects} \label{appx:Res_test}

\begin{figure}
    \centering
    \includegraphics[width = \columnwidth]{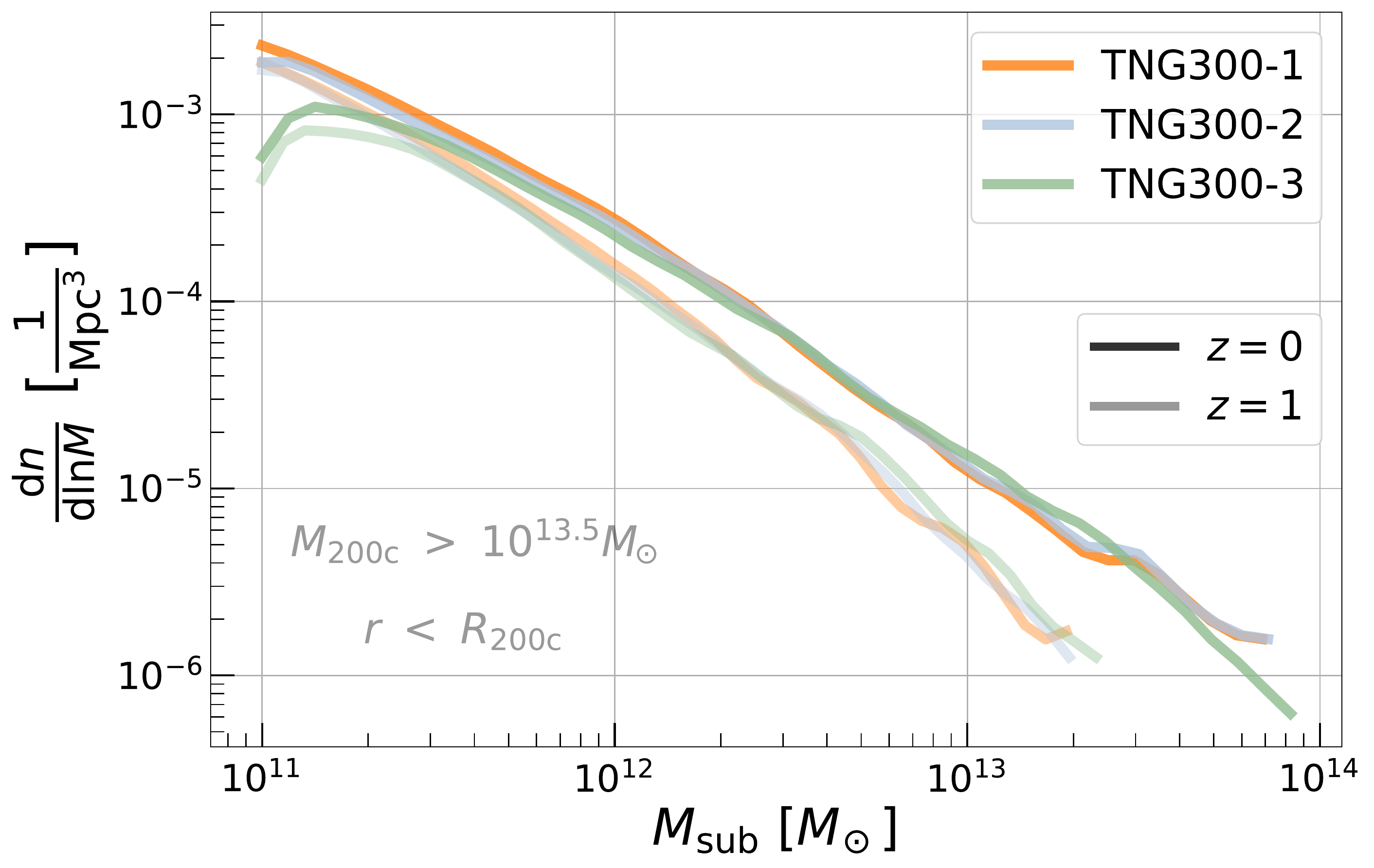}
    \caption{The conditional satellite subhalo mass function for TNG300 runs of different resolution (colors) for two redshifts (tones). We only show subhaloes within $\Rtwohc$ of host haloes with $\Mtwohc > 10^{13.5} \msol$. The two higher resolution runs converge to the same answer across the whole mass range shown here, whereas the lowest resolution run begins diverging below $M_{\rm sub} < 10^{11.5} \msol$.}
    \label{fig:Subhalo_Mass_Function}
\end{figure}

\begin{figure}
    \centering
    \includegraphics[width = \columnwidth]{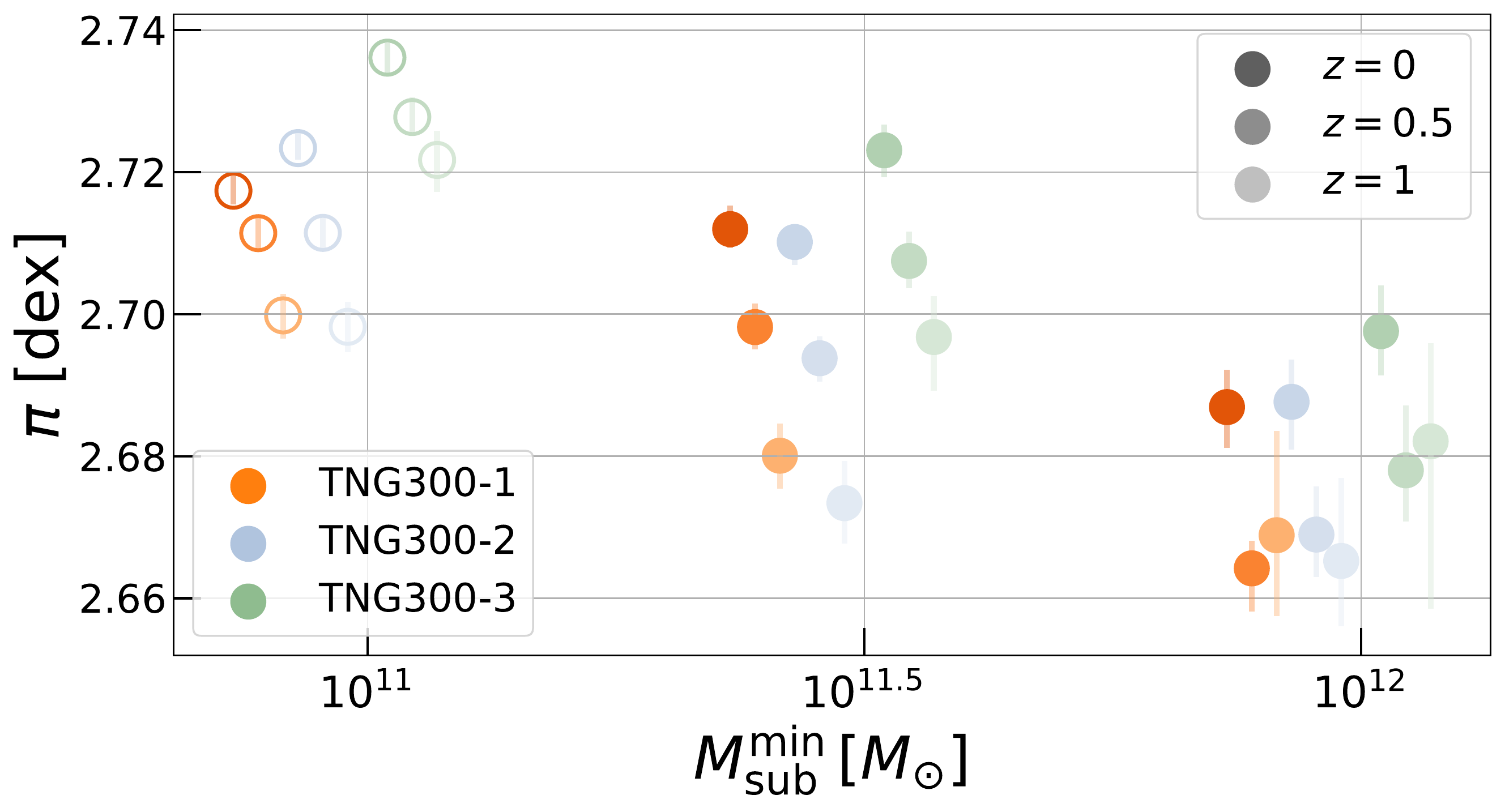}
    \caption{The normalization of the satellite subhalo velocity dispersion scaling with host halo mass, for different subhalo mass thresholds (denoted here by $M_{\rm sub}^{\rm min}$), and for the three different resolutions of the TNG300 suite. The normalizations of the TNG300-1 and TNG300-2 runs are quite similar, whereas the TNG300-3 run has a consistently larger normalization. The results for $M_{\rm sub} > 10^{11} \msol$ are shown as open symbols to highlight that the different resolution runs have significant divergence in the satellite subhalo mass function at this subhalo mass, as shown in Figure \ref{fig:Subhalo_Mass_Function}.}
    \label{fig:Subhalos_Params_ResTest}
\end{figure}

The simulations we consider in our analysis all have different astrophysical model prescriptions, but they also have different resolutions, and it is difficult to fully disentangle how much each effect contributes to the overall differences we observe between simulations (eg. Figure~\ref{fig:Param_evolution}). To shed light on the impact of resolution, we use the \textsc{TNG300} suite, which has three different resolution runs --- TNG300-1, TNG300-2, and TNG300-3 --- to test the resolution dependence of our results. Conveniently, the resolutions of TNG300-2 and TNG300-3 are approximately that of MGTM and BM, respectively. Note that TNG300-1 is the fiducial run that we have used throughout our main analysis.

For this analysis, we focus only on the subhaloes, not galaxies. The stellar mass in galaxies is lower in the lower resolution runs \citep{Pillepich2018FirstGalaxies} because the TNG astrophysical model parameters are fixed for all runs and are not re-tuned for each resolution level \citep{Pillepich2018Methods}. Thus, the three TNG300 runs have different S-GSMFs, and a resolution test that uses galaxy properties would be affected by the absence of the recalibration step. As an alternative, we use the subhalo mass, and limit this study to masses above which the satellite subhalo mass function of the three TNG300 runs, shown in Figure~\ref{fig:Subhalo_Mass_Function}, has approximately converged.

Upon performing EVL for subhaloes, we find clear evidence that the normalizations increase with decreasing resolution (Figure~\ref{fig:Subhalos_Params_ResTest}). The TNG300-1 and TNG300-2 runs show consistency amongst one another for all redshifts, but TNG300-3 differs significantly from these two. This is particularly interesting, as it mirrors our main result where MGTM and TNG300 show similar normalizations, but BM has a significantly higher one. Given that TNG300-2 and TNG300-3 share similar resolution scales with MGTM and BM, respectively, it is possible --- though not necessary --- that a significant part of the discrepancy in BM arises from just resolution differences. The slopes and scatters of the different runs (not shown here) are statistically consistent with one another.

Note that the velocity bias is a ratio between the galaxy and DM velocity dispersions and thus will be unaffected by resolution \textit{as long as} both dispersions are similarly biased by resolution effects. However, in this section, we are not studying the impact of resolution on $b_v$, but rather pointing out that simulations that exist in the high-normalization end of our $\sigmaSat$ results (Figure~\ref{fig:Param_evolution}) may suffer from potential resolution issues.

\section{Additional Model Tests} \label{appx:likelihood_tests}

Here we detail two additional tests of the EVL model. We have assumed throughout our main analysis that our scaling parameters do not run with halo mass, and we validate this assumption in section \ref{appx:Mass_dependence}. Next, in section \ref{appx:Subsampling}, we demonstrate the power of our likelihood method in obtaining constraints even from very sparsely populated data

\subsection{Mass-dependence of galaxy EVL scaling parameters}\label{appx:Mass_dependence}

\begin{figure} \label{fig:Mass_dependence}
    \centering
    \includegraphics[width = \columnwidth]{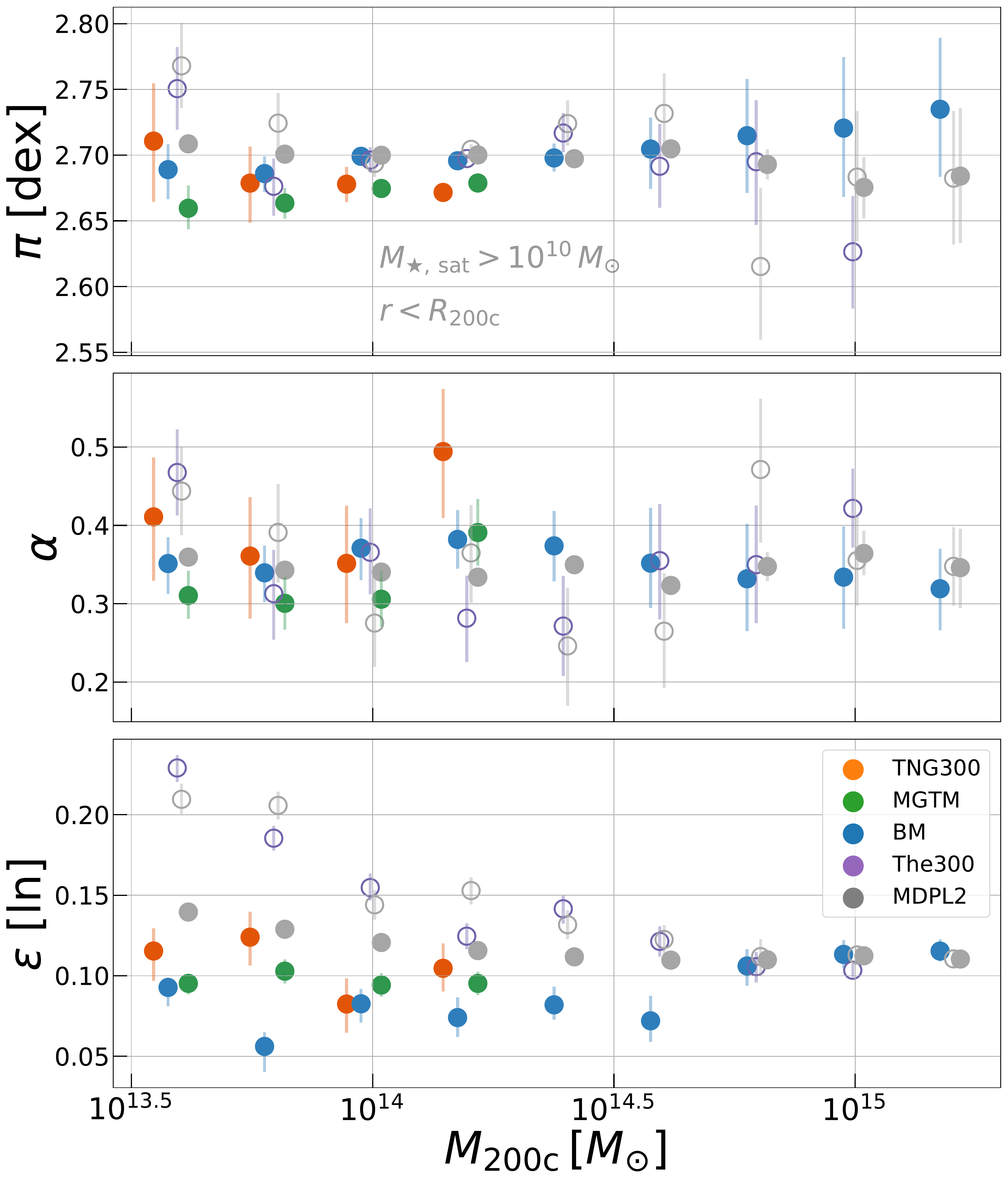}
    \caption{Evolution of the scaling parameters as a function of \textit{halo} mass. We choose bins of width $\Delta M = 0.2 \,\,{\rm dex}$ centered on an $\Mtwohc$ mass scale and compute the parameters using only haloes in those bins. Cosmologically mass-complete (mass-incomplete) simulations are shown with closed (open) circles. \textsc{MDPL2} has both the original mass-complete sample, and a modified sample that mimics the mass-incompleteness resulting from the zoom-in nature of The300 simulations. The normalizations and slopes are statistically consistent with no mass-dependence. For the scatter, however, \textsc{The300} shows a strong dependence on halo mass, and the ``incomplete'' \textsc{MDPL2} sample captures this behavior very well.}
\end{figure}

In our likelihood model, we assume that the slope $\alpha$ and scatter $\epsilon$ are independent of host halo mass $\Mtwohc$. We explicitly test this assumption by measuring the scaling relation parameters only using host haloes within mass bins of width $0.2$ dex. The results are shown in Figure~\ref{fig:Mass_dependence}, where the normalization is still quoted at a scale of $h(z)\Mtwohc = 10^{14} \msol$. There are two sets of gray symbols (open and closed) --- the closed gray symbols are results from the fiducial MDPL2 sample while the open gray symbols are results from the \textsc{MDPL2} sample that has been modified to replicate the incomplete mass-function of \textsc{The300}. The latter is constructed by selecting all MDPL2 haloes of masses below which The300 sample is incomplete, and then preferentially selecting only those MDPL2 haloes that are within 22 (comoving) $\mpc$ of the larger haloes from the mass-complete part of the MDPL2 halo sample. We have verified the consistency between the mass functions of The300 and the modified MDPL2 sample.

From our analysis we are able to make two claims: (i) TNG300, MGTM, and BM show no clear mass-dependence in the parameters. The variations are either stochastic, or are within the errorbars of the measurements. (ii) The increased scatter in The300 at low halo mass can be mimicked by a similar sample constructed using MDPL2. While we cannot make any robust claim on the \textit{cause} of the increased scatter, this result implies the underlying cause is at least \textit{correlated} to a selection effect on the local environment of the low-mass haloes.

The slopes and normalizations shown in Figure~\ref{fig:Mass_dependence} are all statistically consistent with one another, with no preference for any mass evolution. Note that this is the case even for the modified \textsc{MDPL2} sample. We take this as validation that \textsc{The300} sample's mean $\sigmaSat - \Mtwohc$ relation is quite insensitive to the environment-based sample selection for low-mass haloes, whereas the scatter is clearly impacted by this. However, given the velocity bias only depends on the mean relation --- and not the scatter --- we continue using the entire mass range of \textsc{The300} in our main analysis.

\subsection{Sparse sampling} \label{appx:Subsampling}

\begin{figure}
    \centering
    \includegraphics[width = \columnwidth]{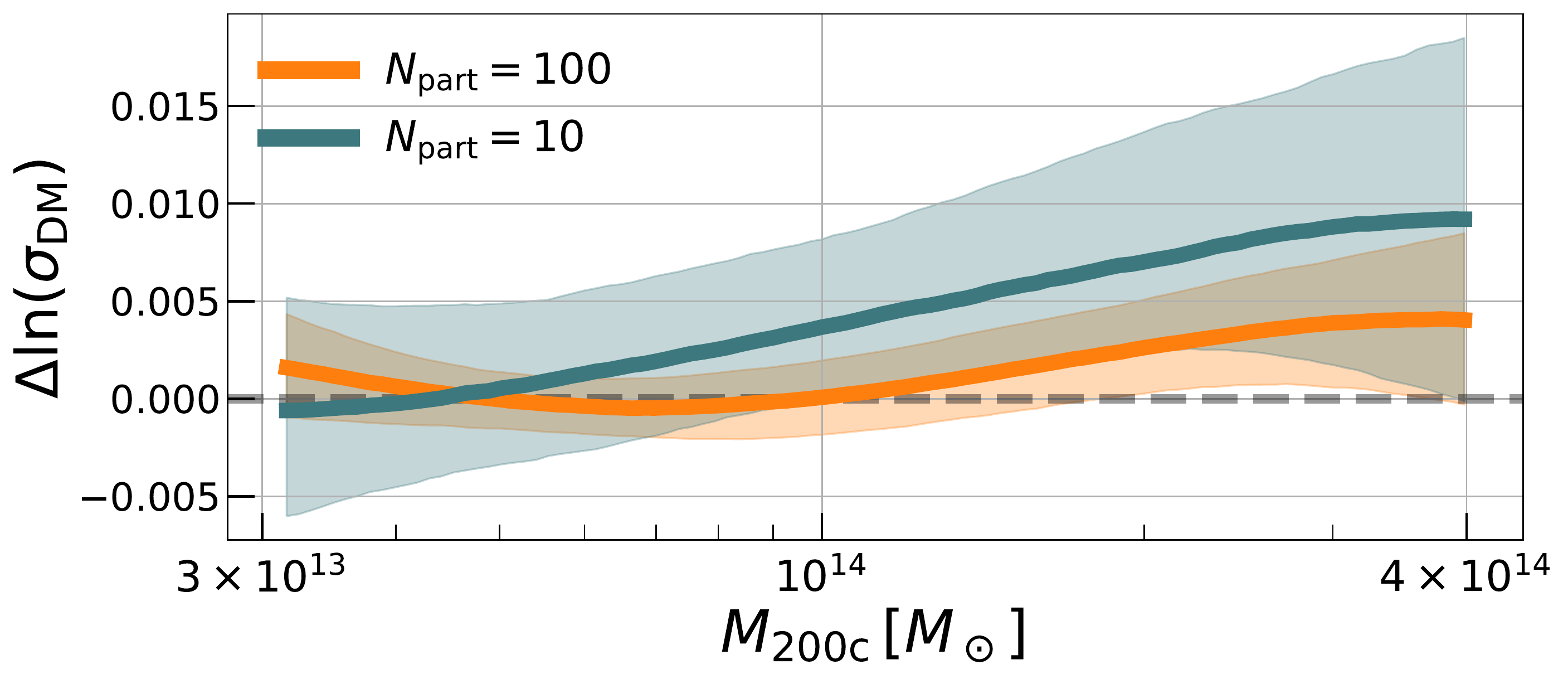}
    \caption{The fractional difference between the EVL-derived and KLLR-derived estimate of the $\sigmaDM - \Mtwohc$ relation. For the likelihood, we present two versions --- one where we use $N_{\rm part} = 100$ particles, and another where we use only $N_{\rm part} = 10$ particles.}
    \label{fig:Subsampling}
\end{figure}

In our main analysis, we showed that the likelihood estimator provides result consistent with those from commonly used regression methods, in the limit $N_{\rm part} \gg 1$ (Figure \ref{fig:Method_comparisons}). Here, we test how well the likelihood method does when we downsample the dataset. We perform the exact same analysis as in Figure~\ref{fig:Method_comparisons}, but instead of using $N_{\rm part} = 100$ DM particles, we use only $N_{\rm part} = 10$ particles. We also focus on the mean relation, $\sigmaDM - \Mtwohc$ instead of its slope and scatter.

We find that the results from the likelihood estimator are statistically consistent with those from \textsc{Kllr} (Figure~\ref{fig:Subsampling}). Even the upper bounds of the $N_{\rm part} = 10$ result are at most within $1.5\%$ of the \textsc{Kllr} estimate.

\section{Sensitivity to Aperture} \label{appx:Aperture}

\begin{figure*}
    \centering
    \includegraphics[width = 2\columnwidth]{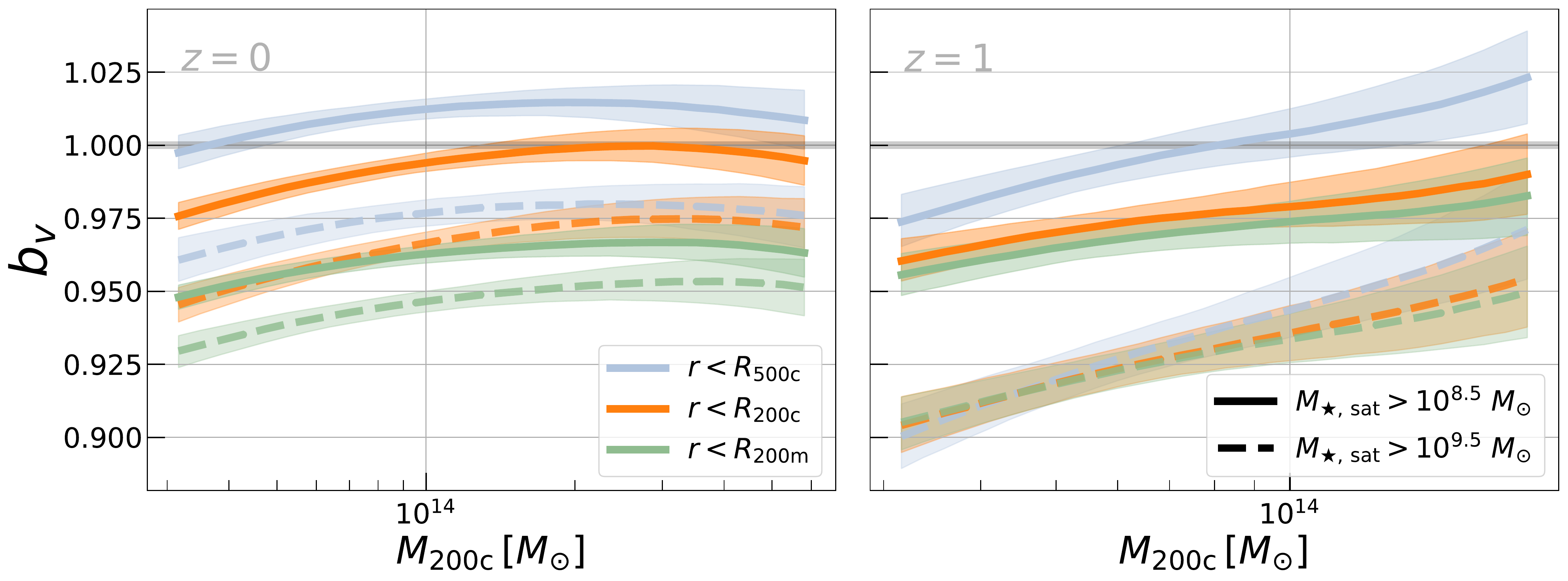}
    \caption{Similar to Figure~\ref{fig:Velocity_Bias}, but for a single \textsc{TNG300} sample with different aperture choices. Smaller apertures lead to a higher $b_v$, and the relative differences between the $b_v$ of different apertures can depend on all of $\Mstarsat$, $\Mtwohc$ and $z$.}
    \label{fig:Velocity_Bias_Aperture}
\end{figure*}

For our main analysis, we have computed the velocity dispersion (both for galaxies and DM) with an aperture defined by the radius $\Rtwohc$. While this is a common choice for observational work on galaxy velocity dispersions \citep[\eg][F16]{Sifon2016VelDispACT}, it is still a potential free parameter in future analyses. Here, we redo our analysis of TNG300 but replace this aperture choice with two other commonly used radii for cluster-scale haloes, $\Rfivehc$ and $\Rtwohm$. Both differ from $\Rtwohc$ in only the density contrast used in the radius definition, with $\Rfivehc$ using $\rho_\Delta = 500 \rho_c(z)$, and $\Rtwohm$ using $\rho_\Delta = 200\rho_m(z)$. Note that our mass variable continues to be $\Mtwohc$, as we are not studying the impact of changing the spherical overdensity definition, but only of changing the satellite galaxy (and DM particle) sample.

We find that using smaller apertures leads to a higher velocity bias (Figure \ref{fig:Velocity_Bias_Aperture}), and this trend was noted in previous studies of both observations \citep{Sifon2016VelDispACT} and simulations \citep{Lau2010BaryonDissipationVelDisp, Armitage2018CEagleVelBias, Ferragamo2020VelDispEstimators}. We also find that the relative difference between the $b_v$ from different apertures can depend on $\Mstarsat$, $\Mtwohc$ and $z$. This dependence arises from changes in the normalization and slope of the $\sigmaSat-\Mtwohc$ relation due to varying the aperture of the measurements. Note also that at high redshift, where we have $\rho_m(z) \approx \rho_c(z)$ and so $\Rtwohm \approx \Rtwohc$, the bias computed within both radii are statistically consistent as expected.

% Don't change these lines
\bsp	% typesetting comment
\label{lastpage}
\end{document}